\documentclass[longauth]{aa}  
\usepackage{graphicx}
\usepackage{txfonts}
\usepackage{lipsum}
\usepackage{subcaption}         
\usepackage{placeins}       
\usepackage{tabularx}
\usepackage{pdflscape}

\usepackage{xcolor}
\usepackage[urlcolor=cyan, colorlinks=true, citecolor=blue, linkcolor=blue]{hyperref}
\usepackage{siunitx}

\begin{document}

\title{Time-Domain Photometry and Activity Evolution of Interstellar Comet 3I/ATLAS with BHTOM}


\author{
A.~Fraser~Gillan\inst{\ref{ncbj}}
\and Łukasz~Wyrzykowski\inst{\ref{ncbj},\ref{uowarsaw},\ref{easst}}
\and Przemys{\l}aw~J.~Miko{\l}ajczyk\inst{\ref{wroclaw},\ref{ncbj}} 
\and Krzysztof~Kotysz\inst{\ref{wroclaw},\ref{uowarsaw}} 
\and Erica~Bufanda\inst{\ref{edinburgh}} 
\and Colin~O.~Chandler\inst{\ref{UW},\ref{LSST_tucson},\ref{NAU}} 
\and Süleyman~Fişek\inst{\ref{istanbul_ast},\ref{istanbul_obs}} 
\and Henry~H.~Hsieh\inst{\ref{PSI}} 
\and Michael~S.~P.~Kelley\inst{\ref{UMD}} 
\and Priscila~J.~Pessi\inst{\ref{ncbj}} 
\and James~E.~Robinson\inst{\ref{edinburgh}} 
\and Sinan~Aliş\inst{\ref{istanbul_ast},\ref{istanbul_obs}} 
\and Wie{\'n}czys{\l}aw~Bykowski\inst{\ref{mps},\ref{easst}}
\and Richard~E.~Cannon\inst{\ref{edinburgh}} 
\and Martin~Dominik\inst{\ref{standrews}} 
\and Barbara~Handzlik\inst{\ref{Jagiellonian}} 
\and Mehmet~İçen\inst{\ref{istanbul_ast}} 
\and Sebastian~Kurowski\inst{\ref{Jagiellonian}} 
\and Ahmet~Cem~Kutluay\inst{\ref{keele},\ref{ankara},\ref{kreiken}} 
\and Joysankar~Majumdar\inst{\ref{uowarsaw}} 
\and Çağlayan~Nehir\inst{\ref{tug}} 
\and David~O'Neill\inst{\ref{birmingham}} 
\and Sibel~Ötken\inst{\ref{istanbul_ast}} 
\and Kangming~Pu\inst{\ref{monash}} 
\and Özlem~Şimşir\inst{\ref{ankara},\ref{kreiken}} 
\and Colin~Snodgrass\inst{\ref{edinburgh}} 
\and Cihan~Tuğrul~Tezcan\inst{\ref{dag}} 
\and Fatma~Tezcan\inst{\ref{atatürk}} 
\and Mauritz~Wicker\inst{\ref{uowarsaw},\ref{inglapalma}} 
\and Fuat~Korhan~Yelkenci\inst{\ref{istanbul_ast}} 
\and Michał~Żejmo\inst{\ref{Zielona_Gora}}
\and Kendall~Ackley\inst{\ref{warwick}} 
\and M.~Andersen\inst{\ref{copenhagen}} 
\and C.~Ávalos-Vega\inst{\ref{antofagasta}} 
\and Sergey~Belkin\inst{\ref{monash}} 
\and V.~Bozza\inst{\ref{salerno},\ref{infn_napoli}} 
\and Rene~P.~Breton\inst{\ref{manchester}} 
\and M.~J.~Burgdorf\inst{\ref{hamburg}} 
\and Jorge~Casares\inst{\ref{iac}} 
\and Vik~Dhillon\inst{\ref{sheffield}} 
\and A.~Donaldson\inst{\ref{edinburgh}} 
\and Martin~J.~Dyer\inst{\ref{sheffield}} 
\and R.~Figuera~Jaimes\inst{\ref{atacama},\ref{mas},\ref{puc_chile},\ref{standrews}} 
\and Duncan~K.~Galloway\inst{\ref{monash}} 
\and T.~C.~Hinse\inst{\ref{sdu}} 
\and M.~Hundertmark\inst{\ref{ari_heidelberg}} 
\and E.~Khalouei\inst{\ref{snu}} 
\and Thomas~Killestein\inst{\ref{warwick}} 
\and Rubina~Kotak\inst{\ref{turku}} 
\and Amit~Kumar\inst{\ref{warwick}} 
\and Feng-Yuan~Frey~Liu\inst{\ref{edinburgh}} 
\and P.~Longa-Pe{\~n}a\inst{\ref{antofagasta}} 
\and Joe~Lyman\inst{\ref{warwick}} 
\and Luigi~Mancini\inst{\ref{inaf_torino},\ref{torvergata}} 
\and A.~Moharana\inst{\ref{keele}} 
\and V.~Molina\inst{\ref{atacama}} 
\and Kanthanakorn~Noysena\inst{\ref{narit}} 
\and Laura~Kate~Nuttall\inst{\ref{portsmouth}} 
\and Paul~O'Brien\inst{\ref{leicester}} 
\and V.~Okoth\inst{\ref{edinburgh}} 
\and C.~Opitom\inst{\ref{edinburgh}} 
\and Don~Pollacco\inst{\ref{warwick}} 
\and M.~Rabus\inst{\ref{ucssc}} 
\and Gavin~Ramsay\inst{\ref{armagh}} 
\and S.~Sajadian\inst{\ref{perimeter}} 
\and A.~Salinas~San~Martin\inst{\ref{atacama}} 
\and J.~Skottfelt\inst{\ref{openu}} 
\and J.~Southworth\inst{\ref{keele}} 
\and Danny~Steeghs\inst{\ref{warwick}}
\and J.~Tregloan-Reed\inst{\ref{antofagasta}} 
\and Krzysztof~Ulaczyk\inst{\ref{warwick}} 
\and R.~Vieliute\inst{\ref{standrews},\ref{inglapalma}} 
}

\institute{
Astrophysics Division, National Centre for Nuclear Research, Pasteura 7, 02-093 Warsaw, Poland \label{ncbj}
\and Astronomical Observatory, University of Warsaw, Al.~Ujazdowskie~4, 00-478 Warsaw, Poland \label{uowarsaw}
\and European Astronomical Society of Small Telescopes (EASST.eu) \label{easst}
\and Astronomical Institute, University of Wrocław, ul.~Mikołaja~Kopernika~11, 51-622 Wrocław, Poland \label{wroclaw}
\and Institute for Astronomy, University of Edinburgh, Royal Observatory, Edinburgh, EH9~3HJ, UK \label{edinburgh}
\and Dept. of Astronomy \& the DiRAC Institute, University of Washington, 3910 15th Ave NE, Seattle, WA 98195, USA \label{UW}
\and LSST Interdisciplinary Network for Collaboration and Computing, 933 N. Cherry Avenue, Tucson, AZ 85721, USA \label{LSST_tucson}
\and Dept. of Astronomy \& Planetary Science, Northern Arizona University, PO Box 6010, Flagstaff, AZ 86011, USA \label{NAU}
\and Department of Astronomy and Space Sciences, Faculty of Science, Istanbul University, 34116 Istanbul, Türkiye \label{istanbul_ast}
\and Istanbul University Observatory Research and Application Center, 34116 Istanbul, Türkiye \label{istanbul_obs}
\and  Planetary Science Institute, 1700 East Fort Lowell Road, Suite 106, Tucson, AZ 85719, USA \label{PSI}
\and Department of Astronomy, University of Maryland, College Park, MD 20742-0001, USA \label{UMD}
\and Max Planck Institute for Solar System Research, 37077 Göttingen, Germany \label{mps}
\and Centre for Exoplanet Science, SUPA School of Physics \& Astronomy, University of St~Andrews, North~Haugh, St~Andrews KY16~9SS, UK \label{standrews}
\and Astronomical Observatory, Jagiellonian University, ul. Orla 171, 30-244, Kraków, Poland \label{Jagiellonian}
\and Astrophysics Group, Keele University, Staffordshire, ST5 5BG, UK \label{keele}
\and Ankara University, Faculty of Science, Astronomy \& Space Sciences Department, Tandogan, TR-06100 Ankara, Türkiye \label{ankara}
\and Ankara University, Astronomy and Space Sciences Research and Application Center (Kreiken Observatory), İncek Blvd., TR-06837, Ahlatlıbel, Ankara, Türkiye \label{kreiken}
\and Türkiye National Observatories, TUG, 07070 Antalya, Türkiye \label{tug}
\and School of Physics and Astronomy, University of Birmingham, Birmingham B15~2TT, UK \label{birmingham}
\and School of Physics and Astronomy, Monash University, Clayton, VIC~3800, Australia \label{monash}
\and Türkiye National Observatories, DAG, 25050, Erzurum, Türkiye \label{dag}
\and Atatürk University, Institute of Science, Department of Astronomy and Astrophysics, 25240 Erzurum, Türkiye \label{atatürk}
\and Isaac Newton Group of Telescopes, Apartado 321, E-38700 Santa Cruz de La Palma, Spain \label{inglapalma}
\and Janusz Gil Institute of Astronomy, University of Zielona Gora, Szafrana 2, 65-516 Zielona Gora, Poland \label{Zielona_Gora}
\and Department of Physics, University of Warwick, Gibbet Hill Road, Coventry CV4~7AL, UK \label{warwick}
\and Niels Bohr Institute, {\O}ster Voldgade 5, 1350 Copenhagen, Denmark \label{copenhagen}
\and Centro de Astronom{\'{\i}}a, Universidad de Antofagasta, Av.\ Angamos 601, Antofagasta, Chile \label{antofagasta}
\and Dipartimento di Fisica "E.R. Caianiello", Universit{\`a} di Salerno, Via Giovanni Paolo II 132, 84084, Fisciano, Italy \label{salerno}
\and Istituto Nazionale di Fisica Nucleare, Sezione di Napoli, Napoli, Italy\label{infn_napoli}
\and Jodrell Bank Centre for Astrophysics, Department of Physics and Astronomy, The University of Manchester, Manchester M13~9PL, UK \label{manchester}
\and Earth System Sciences, Atmospheric Science, University of Hamburg, Hamburg, Germany. \label{hamburg}
\and Instituto de Astrofísica de Canarias, E-38205 La~Laguna, Tenerife, Spain \label{iac}
\and Department of Physics and Astronomy, University of Sheffield, Sheffield S3~7RH, UK \label{sheffield}
\and Instituto de Astronomia y Ciencias Planetarias, Universidad de Atacama, Copayapu 485, Copiapo, Chile \label{atacama}
\and Millennium Institute of Astrophysics MAS, Nuncio Monsenor Sotero Sanz 100, Of. 104, Providencia, Santiago, Chile \label{mas}
\and Instituto de Astrofísica, Facultad de Física, Pontificia Universidad Católica de Chile, Av. Vicuña Mackenna 4860, 7820436, Macul, Santiago, Chile \label{puc_chile}
\and  University of Southern Denmark, Department of Physics, Chemistry and Pharmacy, SDU-Galaxy, Campusvej 55, 5230, Odense M, Denmark \label{sdu}
\and Astronomisches Rechen-Institut, Zentrum f{\"u}r Astronomie der Universit{\"a}t Heidelberg (ZAH), 69120 Heidelberg, Germany \label{ari_heidelberg}
\and Department of Physics, Ewha Womans University, 52 Ewhayeodae-gil,
Seodaemun-gu, Seoul 03760, Republic of Korea \label{snu}
\and Department of Physics \& Astronomy, University of Turku, Vesilinnantie~5, FI-20014 Turku, Finland \label{turku}
\and INAF -- Turin Astrophysical Observatory, via Osservatorio 20, 10025 Pino Torinese, Italy \label{inaf_torino}
\and Department of Physics, University of Rome ``Tor Vergata'', Via della Ricerca Scientifica 1, 00133 Rome, Italy. \label{torvergata}
\and National Astronomical Research Institute of Thailand, Chiangmai 50180, Thailand \label{narit}
\and Institute of Cosmology \& Gravitation, University of Portsmouth, Portsmouth PO1~3FX, UK \label{portsmouth}
\and School of Physics \& Astronomy, University of Leicester, University Road, Leicester LE1~7RH, UK \label{leicester}
\and Departamento de Matemática y Física Aplicadas, Facultad de Ingeniería, Universidad Católica de la Santísima Concepción, Alonso de Rivera 2850, Concepción, Chile \label{ucssc}
\and Armagh Observatory \& Planetarium, College Hill, Armagh BT61~9DG, UK \label{armagh}
\and Perimeter Institute for Theoretical Physics, 31 Caroline St N, Waterloo, ON N2L 2Y5, Canada \label{perimeter}
\and Centre for Electronic Imaging, Department of Physical Sciences, The Open University, Milton Keynes, MK7 6AA, UK \label{openu}
}


  \abstract
   {Time-domain photometric monitoring is essential for characterizing cometary evolution, particularly for rare interstellar objects with limited observing opportunities.}
   {We aimed to characterize the pre-perihelion photometric behavior and dust activity of the interstellar comet 3I/ATLAS, and to test the capability of the Black Hole Target and Observation Manager (BHTOM) platform and telescope network for coordinated high-cadence non-sidereal observations.}
   {We obtained 70 days of time-series photometry of 3I/ATLAS from 2025 July 4 -- September 11 using 16 telescopes and 1554 images. The data were processed and calibrated with the BHTOM pipeline. High-cadence, multi-band imaging was used to measure the rotation period and color evolution, while the dust activity was quantified via $Af\rho$ measurements.}
   {We present a pre-perihelion light curve of 3I/ATLAS from $R_{h} = 3.18 - 2.19$ au, which exhibited a steady increase of $\sim3$ magnitudes with no evidence of anomalous behavior. We measured a rotation period of $P_{\mathrm{rot}} = 15.98 \pm 0.08$ h. The relative dust production increased from $A(0^\circ)f\rho$ $\sim600-1100$ cm, and the upper limit on the dust mass-loss rate increased from $\leq 217$ kg s$^{-1}$ to $\leq 328$ kg s$^{-1}$. We measured an activity index of $n = -1.24 \pm 0.02$, consistent with a well-developed dust coma. The colors were statistically non-changing, with only a weak, non-significant tendency for 3I/ATLAS to become bluer at $3.5 > R_{h} > 2.2$ au.}
   {}
   
               
\maketitle
    
\section{Introduction}\label{sec:intro}
\nolinenumbers
To date, three interstellar objects (ISOs) on unbound, hyperbolic orbits have been discovered. The first, 1I/‘Oumuamua \citep{2017NaturMeech, 2017CBETWilliams}, was identified on 2017 October 19 by the Panoramic Survey Telescope and Rapid Response System (Pan-STARRS) survey \citep{2016arXivChambers}. Notably, 1I/‘Oumuamua exhibited no apparent coma \citep{2017ApJJewitt} but displayed non-gravitational acceleration consistent with comet-like outgassing \citep{2018NaturMicheli}. The second ISO, 2I/Borisov, discovered on 2019 August 30, showed clear cometary activity and spectral properties similar to that of typical solar system comets \citep{2019ApJFitzsimmons, 2019ApJJewitt, 2019A&AOpitom, 2020NatAsGuzik, 2024Fitzsimmons}. The third known ISO, 3I/ATLAS, was discovered on UT 2025 July 1 \citep{2025MPECDenneau} by the Asteroid Terrestrial-impact Last Alert System (ATLAS; \citealp{2018PASPTonry}) survey’s Chilean unit.

Early observations quickly confirmed the cometary nature of 3I/ATLAS. Independent reports of activity were published by \citet{2025ATelAlarcon}, \citet{2025ATelJewitt}, and \citet{2025ATelMinev}, while a series of follow-up studies further characterized the physical properties (see \citealp{2026MNRASFrincke} and references therein for a further detailed overview). Multi-band imaging and spectroscopy revealed that 3I/ATLAS is redder than most solar system comets \citep{2025RNAASBelyakov, 2025MNRASBolin, 2025ATelChampagne, 2025A&AdelaFuenteMarcos, 2025ApJKareta, 2025MNRASOpitom, 2025ApJSeligman, 2025ApJXing, 2025ApJYang}. \citet{2025ApJXing} also detected the first evidence of water activity in 3I/ATLAS at a heliocentric distance of 3.51 au, while \citet{2025ApJCordiner, 2025RNAASLisse} showed a high CO${_2}$ production rate with a high CO${_2}$/H${_2}$O ratio.

\citet{2025arXivChandler} presented results from the 8.4 m Simonyi Survey Telescope at the NSF–DOE Vera C. Rubin Observatory \citep{2019ApJIvezic}, using 97 images obtained in multiple filters. The analysis presented in \citet{2025arXivChandler} provided astrometry, photometry, and morphological characterisation of the comet’s activity between UT 2025 June 21 and UT 2025 July 7. The study measured a lower-limit V-band absolute magnitude of $H_{V} = (13.32\pm0.16)$ mag, and an equivalent upper-limit effective nuclear radius of $r_n\sim(6.6\pm0.5)$ km. \citet{2025ApJJewittHubble} estimated a smaller radius upper limit of $r_n\leq2.8$~km using a fit to the surface brightness profile measured from Hubble Space Telescope observations. Adopting the smaller radius estimate, the nucleus of 3I/ATLAS is larger than both 1I/‘Oumuamua \citep{2017NaturMeech, 2018NatAsFitzsimmons} and 2I/Borisov \citep{2020NatAsGuzik, 2020ApJJewitt}.

The dynamical properties of 3I/ATLAS have also been investigated. Using the \mbox{\={O}tautahi-Oxford} model \citep{2025AJHopkins}, \citet{2025ApJHopkins} ruled out 3I/ATLAS from originating in the same star system as both 1I/‘Oumuamua and 2I/Borisov. From analysis of the velocity, 3I/ATLAS appears to have originated from the Milky Way’s thick disk, making it the first ISO identified from this region.\\

3I/ATLAS reached perihelion on UT 2025 October 29 at a heliocentric distance of $q$ = 1.36 au, according to JPL Horizons \citep{giorginiJPLsOnLineSolar1996}. The ISO entered solar conjunction in mid-October and reappeared in November 2025; ground-based observations were therefore concentrated in the pre-perihelion interval, in the weeks leading up to solar conjunction. This paper presents high-cadence photometric measurements and analysis of 3I/ATLAS prior to solar conjunction using 16 different telescopes, distributed worldwide in both the northern and southern hemispheres. We followed the evolution of the brightness, dust production and color from 2025 July 04, three days after discovery, to 2025 September 11, and simultaneously tested the capabilities of the Black Hole Target and Observation Manager (BHTOM) for non-sidereal objects.

\section{BHTOM and observations} \label{sec:observations}
\subsection{BHTOM} \label{subsec:BHTOM}
Time-domain astronomy is driven by the current wide-field surveys such as ATLAS \citep{2018PASPTonry}, Pan-STARRS \citep{2016arXivChambers} and ZTF \citep{2019PASPBellm}, and has become essential for studying astrophysical phenomena across a wide range of timescales, from rapidly evolving objects to long-term evolution. The Legacy Survey of Space and Time (LSST; \citealp{2019ApJIvezic}) conducted by the Vera C.\ Rubin Observatory will reach much fainter objects than current sky surveys and is expected to revolutionize transient astrophysics, finding thousands of new events every single night and discovering an order of magnitude more objects in each of the solar system's small body reservoirs \citep{2023ApJSSchwamb}. However, these surveys typically follow fixed cadences and revisit sky-fields every few days to weeks, often in different filters, making high-cadence, single-filter monitoring challenging. BHTOM\footnote{\url{https://bhtom.space}} \citep{2024lsstWyrzykowski, 2025RMxACMikolajczyk} provides an effective complement to survey-based data. The system and its associated telescope network were created in 2013 with the main purpose of the follow-up of alerts from ESA's Gaia Space Mission \citep{2021A&AHodgkin}. BHTOM is not an alert broker; rather, it is the missing link between brokers and the telescopes. BHTOM is an open-source platform that employs a uniform, automated pipeline designed to deliver science-ready photometry. It is based on the Target and Observation Manager (TOM) open-source framework developed by the Las Cumbres Observatory \citep{2018SPIEStreet}. BHTOM enables user-defined observation requests for targeted, high-cadence photometric monitoring of individual objects. BHTOM currently accepts requests for targets for long-term or rapid photometric monitoring, such as transients, microlensing events, quasars, variable stars, and extrasolar planets. At the time of writing, the BHTOM telescope network is made up of $\sim170$ telescopes worldwide and includes robotic, manual, amateur, and professional facilities, with mirror diameters ranging from 0.03 m to 2.5 m.

Users can upload reduced data  (bias-, dark-, and flat-field–corrected) from both charge-coupled device (CCD) and complementary metal–oxide–semiconductor (CMOS) cameras to \mbox{BHTOM}, 
enabling observers beyond the professional community, such as amateur astronomers and school pupils, to contribute. All observations are then automatically processed and standardized in BHTOM in order to provide science-ready, calibrated data. Automatic calibration of CCD and CMOS images is performed using a combination of established software packages, including CCDPhot \citep{2024lsstWyrzykowski, 2025RMxACMikolajczyk}, Source-Extractor (SExtractor; \citealp{1996A&ASBertin}), Software for Calibrating AstroMetry and Photometry (SCAMP; \citealp{2006ASPCBertin}), Dominion Astrophysical Observatory Photometry (DAOPHOT II; \citealp{1987PASPStetson}) and WCSTools\footnote{\url{http://tdc-www.harvard.edu/wcstools/}}. Calibration in BHTOM is performed using Gaia Synthetic Photometry (GaiaSP; \citealp{2023A&AGaiaCollaboration}) derived from the Gaia XP low-resolution spectra, in which the Gaia $G$ bandpass (350 -- 1050 nm, \citealp{2018A&AWeiler}) is decomposed and integrated through the desired bandpasses. This provides color-corrected synthetic magnitudes appropriate for point-spread function (PSF) photometry. A PSF model is constructed automatically for each image, and stellar and target fluxes are measured using PSF fitting. The photometric uncertainty assigned to each measurement is taken from the formal PSF-fitting error returned by DAOPHOT II, which reflects the noise properties of the image and the quality of the PSF model fit \citep{1987PASPStetson}. These uncertainties, therefore, capture both photon noise and fitting residuals and provide an appropriate estimate of the measurement precision for each data point.

The astrometric solution is derived using reference catalogs including the U.S. Naval Observatory (USNO) Robotic Astrometric Telescope (URAT-1; \citealp{2015AJZacharias}), the fourth United States Naval Observatory (USNO) CCD Astrograph Catalog (UCAC-4; \citealp{2013AJZacharias}), United States Naval Observatory B1.0 Catalog (USNO-B1.0; \citealp{2003AJMonet}) and Gaia-DR3 \citep{2023A&AGaiaCollaboration}. Photometric calibration is standardized to magnitudes based on GaiaSP \citep{2023A&AGaiaDR3} and the Two Micron All Sky Survey (2MASS; \citealp{2006AJSkrutskie}) reference catalogs. A detailed description of the BHTOM system, including the data processing and automation architecture, is provided in \citet{2025RMxACMikolajczyk}. Finally, BHTOM combines processed input images with all available archival photometry to produce the most complete calibrated light curves possible.

\subsubsection{BHTOM and the solar system}
BHTOM currently supports observation requests for non–solar system targets, as described in Section~\ref{subsec:BHTOM}. Ongoing development of the platform is extending this functionality to additional science cases, including solar system targets. In this context, users will be able to submit observation requests for both comets and asteroids within a defined observing epoch. BHTOM will also query the archival data for past observations of the requested target and coordinate future observing campaigns. In addition to scheduled monitoring requests, BHTOM will support the rapid deployment of the telescope network for follow-up observations in response to alerts from surveys, brokers, or community triggers, without the requirement for telescope proposals. Although this version of BHTOM is not yet publicly available, in this paper, we used observations of 3I/ATLAS to test the capability of the BHTOM observing network and data-processing framework with a fast-moving solar system small body, in support of future non-sidereal observations.

\subsection{Observations}
\begin{figure*}[t]
\centering
{\includegraphics[width=\textwidth]{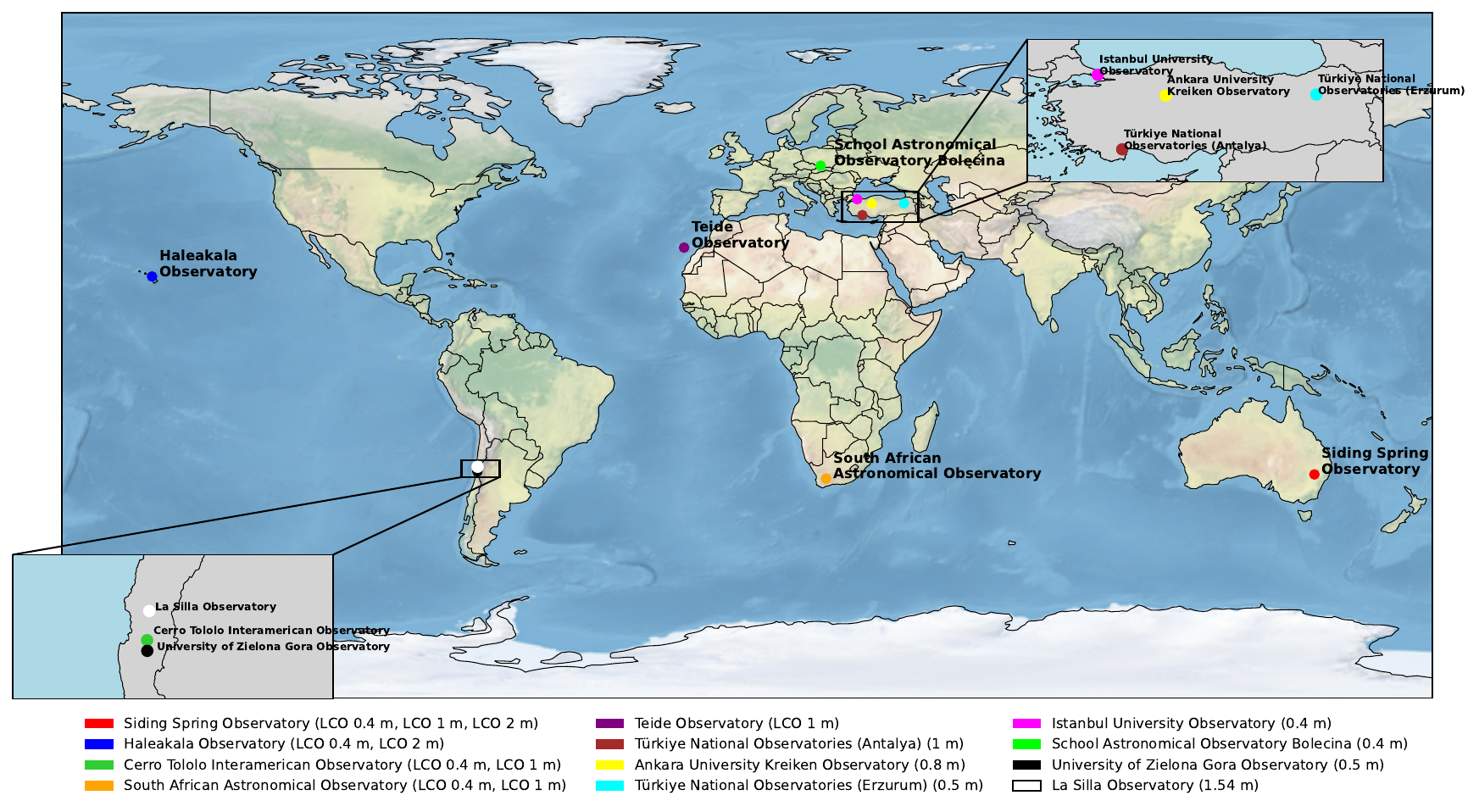}}
	\caption{Geographic locations of the observatories from which photometric data were obtained. Colors correspond to observatory names and telescope apertures given in the legend. Insets indicate regions with clustered sites.}
	\label{fig:observatories_map}
\end{figure*}

The following subsections describe the individual observatories and telescope facilities that contributed data to this study. We used data from both our own allocated observing time with the Las Cumbres Observatory Global Telescope (LCOGT) network facilities, and a variety of observatories that have contributed data using the BHTOM network. The resulting observation log is shown in Table~\ref{tab:observations}, which summarizes all observatories, telescopes, filters, the number of exposures, and the range of observation dates. The dataset combines measurements from both northern and southern hemisphere observatories, providing broad temporal and geographic coverage. All photometry (with the exception of the GOTO data) was performed consistently using the BHTOM photometry pipeline as described in Section~\ref{subsec:BHTOM} to ensure consistency across sites. In this work we used a total of 1554 images for detailed analysis. Figure~\ref{fig:observatories_map} shows the locations of the contributing observatories.

\subsubsection{The LCOGT Network}
The LCOGT network maintains a globally distributed, multi-aperture robotic telescope network \citep{2013PASPBrown}. We obtained data from five observatories in the network, using 0.4~m telescopes at Siding Spring Observatory in Australia, the South African Astronomical Observatory (SAAO), Cerro Tololo Inter-American Observatory (CTIO) in Chile, and Haleakala Observatory in Hawaii, and 1.0~m telescopes at Siding Spring Observatory, SAAO, CTIO, and Teide Observatory in the Canary Islands. The observations were obtained between UT 2025 July 5 and UT 2025 September 11 using both the 0.4~m and 1~m facilities across these sites, see Table~\ref{tab:observations}. The 0.4~m and 1~m telescopes have the same design across all sites, where the 0.4~m telescopes are PlaneWave Delta Rho 350 telescopes equipped with QHY600 CMOS cameras which provide $30' \times 30'$ fields of view (FOVs) with pixel scales of approximately $0\farcs74$/pixel. Meanwhile, the 1~m telescopes are equipped with Sinistro cameras with $13' \times 13'$ FOVs with pixel scales of $0\farcs78$/pixel with $2\times2$ binning in central 2k$\times$2k readout mode. These telescopes used the Sloan $g'$, $r'$, and $i'$ filters with central wavelengths 4770 \AA, 6215 \AA, and 7545 \AA, respectively. All of the observations from the LCOGT network were automatically reduced using the Beautiful Algorithms to Normalize Zillions of Astronomical Images (BANZAI) pipeline \citep{2018SPIEMcCully}. Depending on the telescope and lunar elongation angle, we used 30, 40, or 45 second exposures to minimize trailing effects.

\subsubsection{School Astronomical Observatory Bol\k{e}cina}\label{subsection:SOAB}
The School Astronomical Observatory Bol\k{e}cina (SOAB), located in southern Poland, operates a 0.4 m telescope equipped with a CMOS camera with a pixel scale of \SI{0.238}{\arcsecond}/pixel, providing a field of view of $19.3' \times 12.6'$. Data were collected over three nights: 2025 July 04, 2025 July 20 and 2025 July 23, with a total of 13 images, see Table~\ref{tab:observations}. Observations were obtained using the Johnson-Cousins (hereafter J-C) $R$ and $I$ filters, with individual exposures of 300 seconds for $R$-band and 120 seconds for $I$-band. To mitigate short-timescale scatter and improve the signal-to-noise ratio of the SOAB data, the calibrated photometry was binned in time using 0.5-day intervals. We combined the measurements within each 0.5-day bin in flux space by converting magnitudes to fluxes, computing the weighted-mean flux, and converting back to a magnitude. For the magnitude-flux conversion, we noted that $UBVRI$ are presented in the VEGA system \citep{2023A&AGaiaCollaboration}, and corresponding zero-points were obtained from the Vega stellar spectrum as presented by \citet{2012PASPBessell}. The binning used inverse-variance weighting, with $3\sigma$ clipping applied - only if four or more measurements were present within a bin. The reported uncertainties correspond to the larger of the weighted-mean uncertainty and the intra-bin scatter. 

\subsubsection{Observations with Turkish Telescopes}
We observed 3I/ATLAS using four different telescopes across Türkiye. At the Türkiye National Observatories in Antalya, we used the 1~m TUG100 telescope on UT 2025 July 11. This telescope is equipped with a $4\mathrm{k} \times 4\mathrm{k}$ CCD detector, providing a $21.5' \times 21.5'$ FOV and a pixel scale of $0\farcs31$/pixel. We used the Sloan $r'$, $i'$, and $z'$ filters with $2 \times 2$ on-chip binning and individual exposure times of 60 seconds. These observations were combined in a single-night bin, per filter, to maximize the signal-to-noise ratio. 

We also used the 0.8~m T80 telescope at the Ankara University Kreiken Observatory to observe 3I/ATLAS on UT 2025 July 7. The instrument is equipped with a $1\mathrm{k} \times 1\mathrm{k}$ CCD detector, providing a $12' \times 12'$ FOV with a pixel scale of $0\farcs69$/pixel. Observations were made using the $g'$, $r'$, and $i'$ filters, with $2 \times 2$ on-chip binning. A total of 50 images were taken with exposure times between 30 and 90 seconds. The data were then binned by night, per filter, resulting in three photometric data points. 

We used the 0.5~m ATA050 telescope at the Erzurum site of Türkiye National Observatories on UT 2025 July 7. This telescope is equipped with a QHY268M CMOS camera which provided a $20' \times 14'$ FOV and a pixel scale of $0\farcs19$/pixel. Observations were performed without a photometric filter (i.e. clear) and using $4 \times 4$ on-chip binning with individual exposure times of 90 seconds. A total of nine images were taken and subsequently combined into a single-night bin. The clear-filter measurements were calibrated to the J–C $R$-band using the BHTOM pipeline.

Finally, we used the 0.4~m IST40 telescope at the Istanbul University Observatory on six nights between UT 2025 July 18 and UT 2025 July 27. This telescope has a $3358 \times 2536$ CCD detector, giving a $15.5' \times 11.5'$ field of view and a pixel scale of $0\farcs27$/pixel. These observations were also performed without a photometric filter (clear) with individual exposure times of 90 seconds. All images were acquired using $4 \times 4$ on-chip binning, and the images that were taken were then subsequently binned, per night, to maximize the signal-to-noise ratio. These images were calibrated to the J–C $R$-band using the BHTOM pipeline. All of the images that were binned, similarly to the SOAB data in Section~\ref{subsection:SOAB}, are highlighted in Table~\ref{tab:observations}.

\subsubsection{University of Zielona Góra Observatory}
The University of Zielona Góra Observatory (UZO) hosts a 0.5~m telescope, located at the Deep Sky Chile site in the Rio Hurtado Valley, Chile. We observed 3I/ATLAS on five nights between UT 2025 July 17 and 2025 July 24. The telescope uses a 26 megapixel QHY268M Pro CMOS camera with a pixel scale of $0\farcs23$/pixel and a field of view of \SI{23}{\arcmin} $\times$ \SI{15}{\arcmin}. Each of the observations taken with this telescope were 60~second exposures, taken sequentially in the J-C $B, V$ and $R$ bands, see Figure~\ref{fig:uzo_in_text}. These images were the highest cadence in this study, with a total of 656 images taken over the five nights. 

\subsubsection{The Danish 1.54-metre Telescope}
We used the Danish 1.54 m telescope (D154T) at La Silla Observatory in Chile on three nights between 2025 July 16 and 2025 July 29. The D154T was equipped with the Danish Faint Object Spectrograph and Camera (DFOSC) in imaging mode. DFOSC employs a 2k $\times$ 2k thinned Loral CCD, providing a $13.7' \times 13.7'$ FOV with a pixel scale of $0\farcs4$/pixel \citep{1995MsngrAndersen}. Observations were taken using $B,V,R$ and $I$ filters, using non-sidereal tracking and multiple exposures of 60 seconds to minimize trailing effects, particularly during the first month after discovery when 3I/ATLAS appeared near the Galactic plane in the images.

\subsubsection{GOTO}
The Gravitational-wave Optical Transient Observer (GOTO) project is an array of wide-field optical telescope units located at both the Roque de los Muchachos Observatory on La Palma, and the Siding Spring observatory in Australia. GOTO detected 3I/ATLAS in eight difference images obtained over four nights between 2025 August 17 and 2025 August 23 using the Siding Spring node. Each telescope is equipped with a Finger Lakes Instrumentation ML50100 camera based on the KAF-50100 CCD sensor, which provides a pixel scale of \SI{1.25}{\arcsecond}/pixel and a FOV of $2.1^{\circ} \times 2.8^{\circ}$. Although GOTO was not designed specifically for solar system science, its large field of view enabled observations of 3I/ATLAS, which were obtained using the $L$-band, a broadband filter spanning the Sloan $g'$, $r'$, and Johnson $V$ bands from 400 -- 700 nm \citep{2022MNRASSteeghs}. All images had exposure times of 45 s, with the total number of coadds varying between 2, 3, or 4 depending on the observation. Unlike the other observations presented in this study, the photometry of 3I/ATLAS from GOTO was not processed using the BHTOM pipeline. These data are nevertheless included for completeness, but were obtained using a different processing approach and may therefore exhibit systematic offsets relative to the other datasets used in this study. The AB magnitudes were calibrated against the ATLAS Refcat \citep{2018ApJTonry}, and the $L$-band magnitudes were derived using differential photometry. 

\begin{landscape}
\begin{table}
\caption{This table shows the observatories, telescopes, number of exposures in each filter, the start and end date of observation for individual telescopes in both UT and MJD, and the number of unique nights of observation. $g'$, $r'$, $i'$, and $z'$ correspond to the Sloan filters, $B, V, R$, and $I$ correspond to the J-C filter system, and $L$ corresponds to the GOTO $L$-band. $\dagger$ represents the images that were binned by night.}
\centering
\setlength{\tabcolsep}{3.5pt}
\begin{tabular}{lccccc}
\hline        
Observatory & Telescope & No. observations $\times$ Filter(s) & Range of observation dates & MJD & No. nights\\
\hline
LCOGT-Siding Spring Observatory & 0.4 m & 11 $\times$ $g'$, 36 $\times$ $r'$, 9 $\times$ $i'$ & 2025 July 07 -- 2025 August 27 & 60863 -- 60914 & 14\\
LCOGT-Siding Spring Observatory & 1 m & 5 $\times$ $g'$, 15 $\times$ $r'$, 5 $\times$ $i'$ & 2025 August 12 -- 2025 August 26 & 60899 -- 60913 & 5\\
LCOGT-South African Astronomical Observatory & 0.4 m & 11 $\times$ $g'$, 33 $\times$ $r'$, 11 $\times$ $i'$ & 2025 July 14 -- 2025 September 2 & 60870 -- 60920 & 12\\
LCOGT-South African Astronomical Observatory & 1 m & 13 $\times$ $g'$, 43 $\times$ $r'$, 12 $\times$ $i'$ & 2025 July 19 -- 2025 September 11 & 60875 -- 60929 & 17\\
LCOGT-Teide Observatory & 1 m & 1 $\times$ $g'$, 3 $\times$ $r'$, 1 $\times$ $i'$ & 2025 July 30 & 60886 & 1\\
LCOGT-Cerro Tololo Interamerican Observatory & 0.4 m & 25 $\times$ $g'$, 72 $\times$ $r'$, 19 $\times$ $i'$ & 2025 July 05 -- 2025 September 06 & 60861 -- 60924 & 26\\
LCOGT-Cerro Tololo Interamerican Observatory & 1 m & 17 $\times$ $g'$, 50 $\times$ $r'$, 14 $\times$ $i'$ & 2025 July 14 -- 2025 September 11 & 60870 -- 60929 & 19\\
LCOGT-Haleakala Observatory & 0.4 m & 20 $\times$ $g'$, 53 $\times$ $r'$, 18 $\times$ $i'$ & 2025 July 20 -- 2025 August 27 & 60876 -- 60914 & 20\\
$^{\dagger}$ School Astronomical Observatory Bolęcina (SOAB) & 0.4 m & $8\times R$, $5\times I$ & 2025 July 04 -- 2025 July 23 & 60860 -- 60879 & 3\\
$^{\dagger}$ Türkiye National Observatories & 1 m (TUG100) &  19 $\times$ $r'$, 73 $\times$ $i'$, 26 $\times$ $z'$ & 2025 July 11 & 60867 & 1\\
$^{\dagger}$ Ankara University Kreiken Observatory & 0.8 m (T80) & 4 $\times$ $g'$, 32 $\times$ $r'$, and 14 $\times$ $i'$ & 2025 July 7 & 60863 & 1\\
$^{\dagger}$ Türkiye National Observatories & 0.5 m (ATA050) & 9 $\times$ $R$ & 2025 July 7 & 60863 & 1\\
$\dagger$ Istanbul University Observatory & 0.4 m (IST40) & 70 $\times$ $R$ & 2025 July 18 -- 2025 July 27 & 60874 -- 60883 & 6\\
University of Zielona Góra Observatory (UZO) & 0.5 m & $157\times B$, $249\times V$, $250\times R$ & 2025 July 17 -- 2025 July 24 & 60873 -- 60880 & 5\\
La Silla Observatory & Danish 1.54 m Telescope & $32\times B$, $35\times V$, $35\times R$, $31\times I$ & 2025 July 16 -- 2025 July 29 & 60872 -- 60885 & 3\\
Siding Spring Observatory & 0.4 m (GOTO) & 8 $\times$ $L$ & 2025 August 17 -- 2025 August 23 & 60904--60910 & 4\\
\hline
\end{tabular}
\label{tab:observations}
\end{table}
\end{landscape}

\section{3I/ATLAS measurements}
\subsection{The magnitude evolution of 3I/ATLAS}\label{subsec:brightness_evolution}
\begin{figure*}[t]
\centering
{\includegraphics[width=\textwidth]{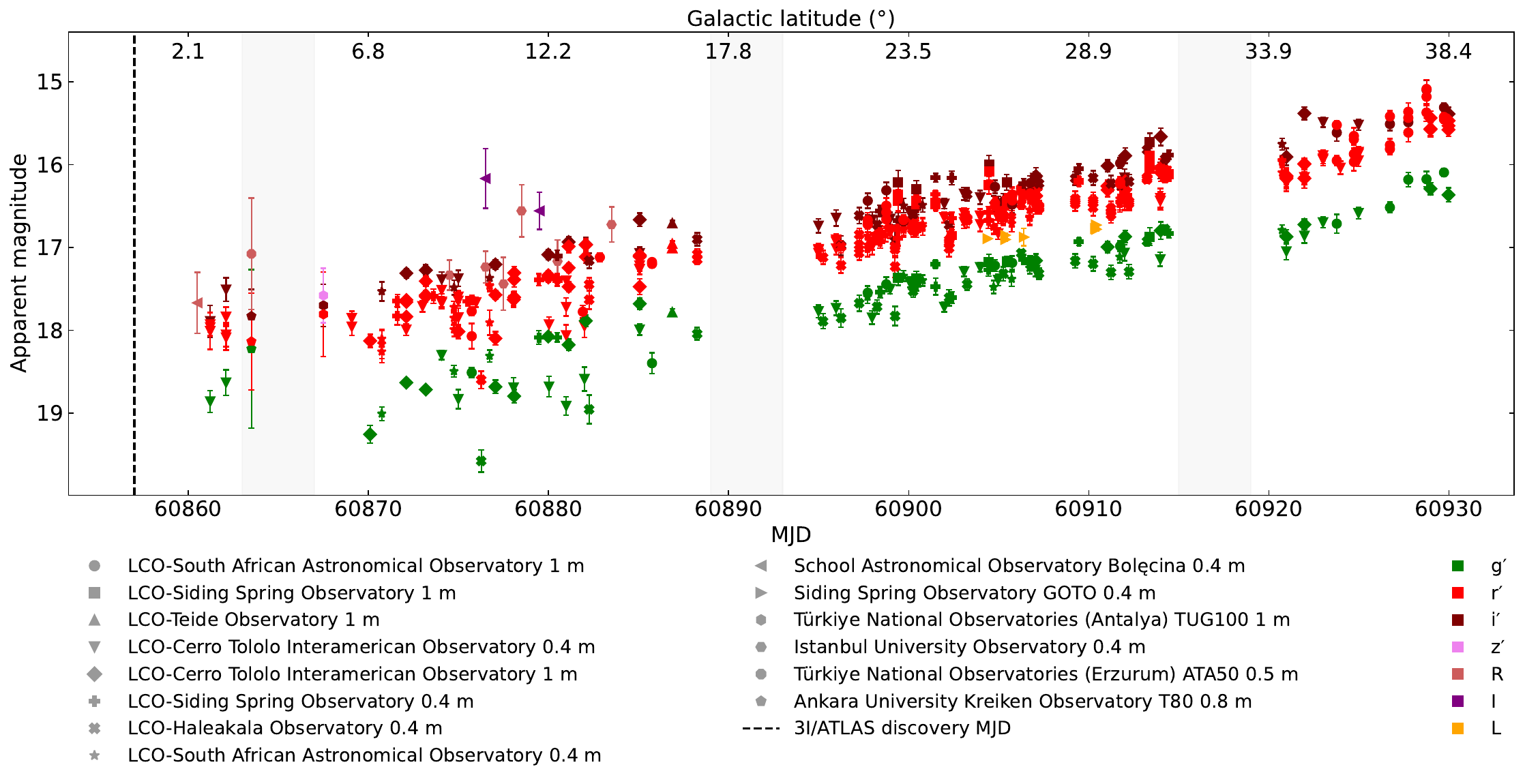}}
     \caption{Apparent magnitude vs MJD of interstellar comet 3I/ATLAS. The calibrated magnitudes are shown as a function of Modified Julian Date, combining observations from multiple telescopes and filters. Marker shapes distinguish individual telescope datasets. $g', r'$ and $i'$ refer to the Sloan filter system and $R$ and $I$ refer to the J-C filter system, and $L$ corresponds to the broad GOTO $L$-band. The vertical dashed line marks the discovery epoch, and the grey shaded regions indicate intervals when the Moon–comet elongation fell below 30\textdegree.}
      \label{fig:master_lightcurve}
\end{figure*}

We measured the apparent magnitude of 3I/ATLAS as a function of Modified Julian Date (MJD) between UT 2025 July 04 and 2025 September 11, using BHTOM photometry calibrated to the GaiaSP system as described in Section~\ref{subsec:BHTOM}. Figure~\ref{fig:master_lightcurve} shows the resulting pre-perihelion light curve, spanning 70 days and combining observations from the facilities listed in Table~\ref{tab:observations}. The dataset comprises high-cadence photometry obtained in Sloan $g'$, $r'$ and $i'$ bands, J-C $R$, and $I$ bands, and the GOTO $L$-band, providing broad multi-band coverage of the comet’s magnitude evolution prior to perihelion. To remove poor-quality measurements, we inspected images corresponding to data points that deviated significantly and unexpectedly from the temporal brightness trend. Sudden increases in brightness are plausible if the comet undergoes an outburst, but inspection showed that the outliers primarily resulted from poor image quality or, more commonly, contamination by nearby background sources where the comet signal became indistinguishable from a star. Such measurements were therefore excluded from the final light curve.

Although PSF photometry suppresses much of the dust-coma signal, it enables magnitude measurements in crowded fields, including the dense stellar regions near the Galactic plane where 3I/ATLAS was located around the time of discovery and during the first months of observations (July–August 2025). In Figure~\ref{fig:master_lightcurve} we indicate the comet’s Galactic latitude over the observing window (top axis), illustrating that 3I/ATLAS moved progressively away from the Galactic plane as the observing campaign progressed. This reduced contamination from nearby background sources and helped to mitigate crowding-related systematics in the later measurements. This approach therefore still provides a useful estimate of the overall brightness and its evolution as the comet moved toward perihelion. However, this approach likely introduced additional uncertainty, as the fraction of coma flux included in the PSF measurement may have varied with seeing conditions and the instrument PSF. In Section~\ref{subsec:dust}, we examine the dust properties of 3I/ATLAS using aperture photometry, which is more appropriate for measuring an extended coma.

Figure~\ref{fig:master_lightcurve} shows that the magnitude of 3I/ATLAS increased steadily over the 70-day observing window, with a total increase of $\sim3$ magnitudes in each of the $g'$, $r'$, and $i'$ bands. We found no evidence in the light curve for behavior that is inconsistent with the monotonic brightening that is typically observed for comets as the heliocentric distance decreases and the corresponding solar insolation increases. Inspecting the images individually, this brightening is consistent with the expansion of the dust coma and corresponding increase in flux. \citet{2025ApJMartinezPalomera} analysed Transiting Exoplanet Survey Satellite (TESS) observations obtained between 2025 May 7 and 2025 June 2, spanning a $\sim26$-day window, and similarly reported a steady brightening with no evidence for anomalous behavior. In Figure~\ref{fig:master_lightcurve}, there are three multi-day gaps present in the dataset which are represented by grey shaded regions. These correspond to time periods where the Moon-comet elongation fell below 30\textdegree, during which observations from the LCOGT were temporarily suspended. We also note that most of the observations at MJD $<$ 60890 (UT 2025 August 3) were obtained with smaller telescopes—primarily the LCOGT 0.4 m units—resulting in lower signal-to-noise ratios and correspondingly larger photometric uncertainties and overall scatter in the data. This was due to the comet brightness being close to the limiting magnitude of the 0.4 m telescopes and the crowded stellar fields.

\subsection{The rotation period}
\begin{figure}[t]
    \centering
    \includegraphics[width=\columnwidth]{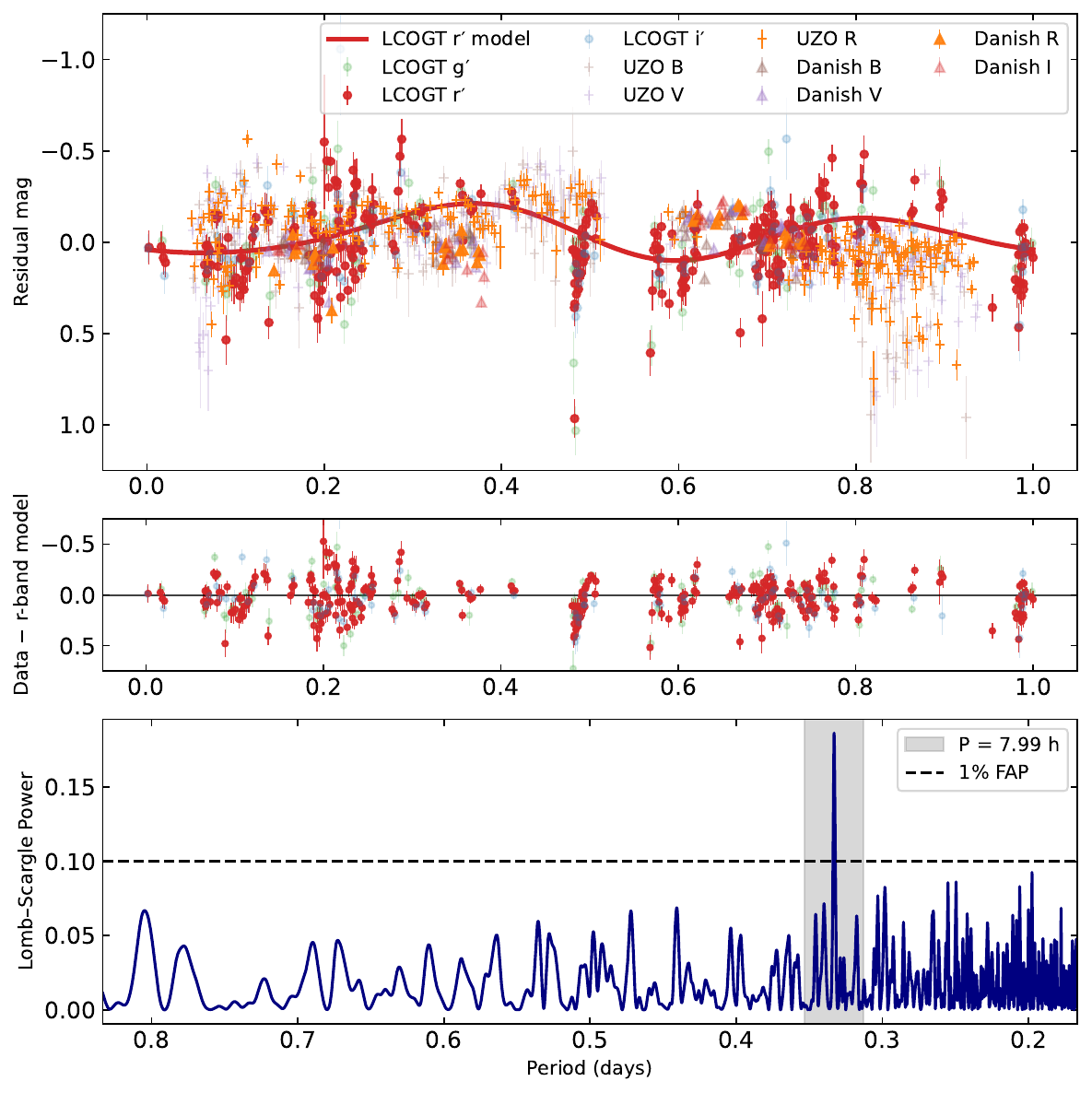}
	\caption{The top panel shows the phase-folded rotational light curve of 3I/ATLAS. Calibrated magnitudes were detrended for the long-term brightening and folded on the adopted rotation period of $P_{\mathrm{rot}} = 15.98 \pm 0.08$ h. LCOGT $r'$, UZO $R$ and D154T $R$ are shown in the foreground with the best-fitting three-harmonic sinusoidal model for the LCOGT $r'$-band over-plotted. The background shows the entire remaining dataset from LCOGT $g'$, $i'$, UZO $B$, $V$ and D154T $B$, $V$ and $I$. The middle panel shows the residuals of the LCOGT data relative to the LCOGT $r'$ harmonic model. The bottom panel shows the LS periodogram of the detrended LCOGT residuals; the shaded region marks the strongest periodicity and the dashed horizontal line indicates the 1\% FAP level.}
	\label{fig:rotation}
\end{figure}

The rotation period, $P_{\mathrm{rot}}$, of 3I/ATLAS was measured using multi-band observations from LCOGT, UZO and the D154T. While the UZO and the D154T datasets provided dense intra-night sampling, they were complementary to the longer-baseline LCOGT observations, which were taken several per night over 70 nights. Inspection of the LCOGT light curve, discussed in Section~\ref{subsec:brightness_evolution}, revealed a monotonic brightening, consistent with the gradual increase in coma size and overall cometary activity. When measuring the rotation period, and therefore focusing on the short-timescale variability associated with nucleus rotation, we de-trended the magnitudes in each of the filters used by the LCOGT by fitting and subtracting a linear function of time. This removed the long-term evolution while preserving variability on hour timescales. The magnitudes measured from the UZO and the D154T datasets were not independently de-trended due to their short temporal baselines. To enable direct comparison between filters and datasets, we computed residual magnitudes by subtracting the mean magnitude of each filter from all measurements in that filter, removing constant offsets. These residuals were then used for the period search and, after a preferred period was identified, the combined residual light curve was phase folded to produce the final rotation light curve, as shown in the top panel of Figure~\ref{fig:rotation}.

A Lomb–Scargle (LS) periodogram \citep{1976Ap&SSLomb, 1982ApJScargle} was computed using the detrended LCOGT residuals, which provided the longest temporal baseline, and the photometric uncertainties were used as the weights in the periodogram calculation. The UZO and D154T datasets were not included in the period search due to their shorter time coverage, which was insufficient to constrain rotation periods on hour timescales. Shown in the bottom panel of Figure~\ref{fig:rotation}, the resulting periodogram exhibited a dominant peak at a period of $P_{\mathrm{LS}} = 7.994$ h, with a false-alarm probability (FAP; the probability that a peak of equal or greater power could arise from noise alone \citealp{2018ApJSVanderPlas}) of $< 1\%$, indicating this peak was unlikely to have arisen by chance. The width of the dominant peak, highlighted in grey, was used to estimate the uncertainty on the detected period, giving an uncertainty of $\pm 0.038$~h. We therefore report a best-fitting photometric periodicity of $P_{\mathrm{LS}} = 7.994 \pm 0.038$~h. 

Interpreting photometric periodicities in an active object was not straightforward as any nucleus-driven signal was likely obscured by the coma, and the light-curve shape could differ substantially from that of an inactive rotating body. In particular, the coma could have diluted or suppressed nucleus-driven light-curve variations due to the changing projected cross-section \citep{1993AJMeech,2009A&AReyniers}. The inner-coma dust brightness can also substantially vary over a single rotation as seen by in-situ measurements of 67P/Churyumov-Gerasimenko \citep{2019A&ATubiana}. If the dust originated from a localized active region on the nucleus, the change in activity during rotation could have mimicked a photometric periodicity. In practice, both a single-peaked signal at $P_{\mathrm{LS}}$ and a double-peaked signal at $2P_{\mathrm{LS}}$ were plausible as shape-driven brightness variations can produce two similar maxima per full rotation \citep{2017MNRASKokotanekova}. \citet{2026A&ASerraRicart} also adopted a doubled-period under the assumption of a jet originating from a single active area near one of the rotational poles of the nucleus. We therefore also considered the doubled interpretation and under this assumption, the LS periodicity corresponded to a nucleus rotation period of $P_{\mathrm{rot}} = 2P_{\mathrm{LS}} = 15.98 \pm 0.08$ h. However, we stress that our photometry alone did not uniquely distinguish between the $P_{\mathrm{LS}}$ and $2P_{\mathrm{LS}}$ solutions, and we therefore treated both as plausible and advise caution when trying to determine periodicity from any coma-obscured nucleus.

The phase-folded LCOGT $g'$, $r'$, and $i'$ residual light curves were then fitted independently using a truncated Fourier series with three harmonics, consisting of a sum of sine and cosine terms plus a constant offset representing the mean residual magnitude in each band. Among the three bands, the $r'$ data exhibited the highest signal-to-noise ratio and the lowest scatter, and were therefore adopted as the reference band. The best-fitting $r'$-band model was evaluated at the phases of all LCOGT, UZO, and D154T data points to determine whether the same phase-dependent behavior was present in the other datasets. The UZO and D154T $R$-band data followed the same phase-dependent behavior as the LCOGT $r'$ data within their photometric uncertainties as shown in the top panel of Figure~\ref{fig:rotation}, indicating that the detected periodicity was consistent across the redder filters used by the different observatories.

The middle panel of Figure~\ref{fig:rotation} shows the $g'$, $r'$ and $i'$ residuals relative to their respective harmonic models, and does not show clear evidence for an additional periodicity beyond the adopted rotation signal. The peak-to-peak amplitude of the light curve was at the level of several tenths of a magnitude; however, for an active object the observed amplitude can be diluted by coma flux and can be sensitive to seeing-dependent variations in the inner-coma photometry \citep{2000AJLicandro}. We therefore did not interpret the measured amplitude in terms of nucleus elongation. The standard deviation of the model-subtracted residuals was $\sigma = 0.19$ mag.

\citet{2025A&AdelaFuenteMarcos} found a rotation period of $P_{\mathrm{rot}} = 16.79 \pm 0.23$ h obtained using the OSIRIS camera spectrograph at the 10.4 m Gran Telescopio Canarias and the Two-meter Twin Telescope (TTT). \citep{2025A&ASantanaRos} found a similar rotation period of $P_{\mathrm{rot}} = 16.16 \pm 0.01$ h using photometric data from multiple telescopes including Telescopi Joan Oró, The Faulkes Telescope South, The Faulkes Telescope North, The Lesedi telescope, and the Nordic Optical Telescope. A month later, in August 2025, \citet{2026A&ASerraRicart} used multiple night imaging from the $TTT$ at the Teide Observatory in the Canary Islands (Spain) to study a faint high-latitude jet in the inner coma. From the jet morphology analysis, the authors determined a rotation period of $P_{\mathrm{rot}} = 15.48 \pm 0.70$ h a month after the two aforementioned studies. The authors attribute this discrepancy to either the intrinsic uncertainties on their methodology, or that 3I/ATLAS has changed its rotation period due to the outgassing torques. Adopting our doubled-period solution of $P_{\mathrm{rot}} = 15.98 \pm 0.08$ h, our measurement was consistent with the value reported by \citet{2026A&ASerraRicart} within $1\sigma$, and consistent with the period reported by \citet{2025A&ASantanaRos} at the $\sim2\sigma$ level. The slightly longer period measured by \citet{2025A&AdelaFuenteMarcos} differed at the $\sim3\sigma$ level. Overall, the published rotation periods for 3I/ATLAS span 15.48--16.79 h, and our measured value of $P_{\mathrm{rot}} = 15.98 \pm 0.08$ h is consistent with this range. As we have previously discussed however, our rotation period should be interpreted with caution as we can not distinguish between the $P_{\mathrm{LS}} = 7.994 \pm 0.038$~h and $P_{\mathrm{rot}} = 15.98 \pm 0.08$~h rotation periods.

\section{Cometary dust activity}
\subsection{Observations, aperture photometry and calibration}\label{subsec:aperture_calibrations}
In the previous sections, we analyzed the evolving magnitude of 3I/ATLAS using PSF photometry calibrated with the BHTOM pipeline. For the analysis of cometary dust activity, however, aperture photometry is more appropriate, as it captures flux from the extended coma rather than assuming a point source, an assumption that does not hold for active comets such as 3I/ATLAS. For optimal aperture photometry, we aimed to minimize the contamination from background sources and avoid over-subtraction of the sky background and therefore we included only images taken when 3I/ATLAS was located in relatively uncrowded stellar fields \citep{2009ARA&AKalberla}. This was away from the Galactic plane at MJD $>60897$ (UT 2025 August 10) at a Galactic latitude of 22\textdegree. Using these data, we measured the apparent magnitude of 3I/ATLAS as a function of time. In total, 83 $r'$-band images obtained with the LCOGT 1 m telescopes were used in the aperture photometry analysis as shown in Table~\ref{ap_phot:observations} which provided the highest signal-to-noise ratios and a sufficiently long baseline.

\begin{table}[ht!]
\caption{Observations used in aperture photometry between UT 2025 August 10 and 2025 September 11 (MJD 60897 and MJD 60929). All telescopes used in aperture photometry were the LCOGT 1 m telescopes.}              
\label{ap_phot:observations}
\centering                        
\begin{tabular}{>{\raggedright\arraybackslash}p{0.72\columnwidth}c}
\hline\hline
Observatory & No. of \\    
& Observations \\
\hline                  
LCOGT Cerro Tololo Inter-American Observatory & 40  \\   
LCOGT Siding Spring Observatory & 12 \\
LCOGT Sutherland Observatory & 31\\
\hline                            
\end{tabular}
\end{table}

The position of 3I/ATLAS in each image was determined by the predicted positions obtained from ephemerides generated using JPL Horizons \citep{giorginiJPLsOnLineSolar1996}. For each image, the predicted sky position at the time of observation was converted to detector coordinates using the image WCS, and the exact centre of 3I/ATLAS was further refined via centroiding in order to ensure accurate aperture placement. The centroid refinement applied pixel-coordinate corrections relative to the ephemeris-predicted location of order $\Delta x, \Delta y \sim (0.4 - 3.9)$ pixels, corresponding to $\sim 0.3\arcsec - 2.9\arcsec$. The centroid was measured by fitting a two-dimensional Gaussian profile around the predicted position using the \emph{Photutils centroid$\_$2dg}\footnote{\url{https://photutils.readthedocs.io/en/stable/api/photutils.centroids.centroid_2dg.html}} routine. To measure the local sky background around the comet, we measured the flux within a circular sky annulus at $5\times10^{4}$ km from the comet. This distance was selected in order to be sufficiently close to the comet where the sky was representative of the flux within the aperture, but far away enough that we avoided any coma contamination within the annulus during sky subtraction. A fixed circular aperture corresponding to a radius of 10,000 km at the comet was adopted for all images. This radius is typically employed for active comets \citep{2013A&ASnodgrass, 2022MNRASGardener, 2024PSJGillan, 2025PSJGillan, 2025arXivChandler, 2025MNRASBolin} in order to ensure changes in the measured flux reflect the intrinsic evolution of 3I/ATLAS within a constant physical aperture, while also facilitating direct comparison with other studies.

Stellar centroids were measured, and instrumental magnitudes were derived using circular apertures with radii defined relative to each measured stellar FWHM, with local sky backgrounds estimated from the surrounding annuli. Stars with poorly defined or non-stellar radial profiles, including saturated sources and objects affected by blending or edge effects, were excluded from the calibration sample. This ensured that only well-behaved stellar profiles were used to determine the photometric solution. Standard aperture corrections and zero-point corrections were then applied to the instrumental magnitudes. Photometric calibration was performed independently for each image using field stars returned from the \emph{Gaia Data Release 3} catalog \citep{2016A&AGaia, 2023A&AGaiaDR3} via the \emph{Astroquery} \citep{2019AJAstroquery_Ginsburg} \emph{cone\_search} routine centred on the location of 3I/ATLAS. We searched within a \SI{290}{\arcsecond} radius. The Gaia $G$-band (350-1050 nm; \citealp{2018A&AWeiler}) is an unfiltered wide band (white) and the integrated flux of the low-resolution blue photometer (BP) and red photometer (RP) spectra give $G_{BP}$ and $G_{RP}$ magnitudes from 350 -- 680 nm and 630 -- 1050 nm respectively \citep{2010A&AJordi, 2018A&AWeiler}. Gaia $G$-band magnitudes and $(BP-RP)$ colors from the catalog were converted to Sloan $g'$ and $r'$ magnitudes using published color transformations\footnote{\url{https://gea.esac.esa.int/archive/documentation/GDR3/Data_processing/chap_cu5pho/cu5pho_sec_photSystem/cu5pho_ssec_photRelations.html}} to give the calibrated $r'$-band magnitudes that were used.

When the calibrated magnitudes were calculated, we manually inspected each outlying measurement from the resulting light curve. We rejected any magnitude measurements of 3I/ATLAS that were affected by cloud cover, had poor image quality, close proximity to a nearby bright object resulting in significant flux contamination within the aperture, or cases where the comet lay several pixels from the edge of the frame which prevented a reliable background measurement. The resulting apparent magnitude of 3I/ATLAS, measured within a 10,000 km radius aperture, as a function of time is shown in the top panel of Figure~\ref{fig:aperture_dust_lc}.

\subsection{The dust production rate}\label{subsec:dust}
In order to quantify the relative dust activity from 3I/ATLAS, we used the $Af\rho$ parameter which is commonly used as a proxy for the dust production rate in comets \citep{1984AJAHearn}. $Af\rho$ was measured from the calibrated aperture magnitudes using Equation~\ref{eq:AfpMag}. 

\begin{equation} \label{eq:AfpMag}
Af\rho = \frac{4 R_{h}^{2} \Delta^{2}}{\rho} \cdot 10^{0.4(m_{\odot} - m_{c})}
\end{equation}

\begin{figure}
    \centering
    \includegraphics[width=\columnwidth]{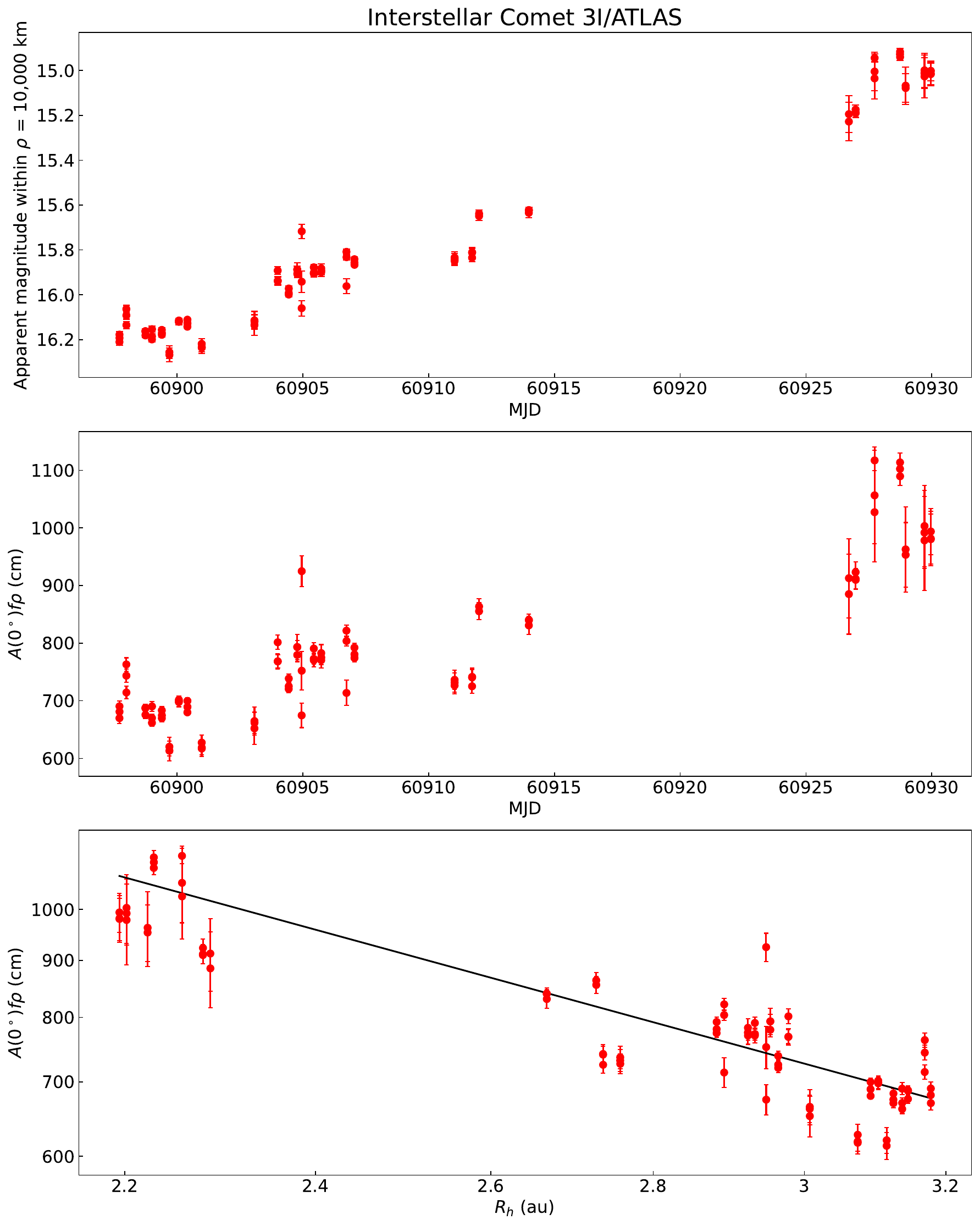}
	\caption{The aperture photometry light curve of 3I/ATLAS is displayed in the top panel and the dust production parameter $A(0^\circ)f\rho$ vs time in MJD is displayed in the middle panel. The lower panel shows $A(0^\circ)f\rho$ vs heliocentric distance, with the solid line representing the $A(0^\circ)f\rho$ activity index, $n$. The red markers on the plots represent the 1 m LCOGT $r$-band measurements.}
	\label{fig:aperture_dust_lc}
\end{figure}

Here, $m_{\odot}$ and $m_{c}$ are the magnitude of the Sun and 3I/ATLAS, respectively, in the same filter. $R_{h}$ is the heliocentric distance in au, and $\Delta$ is the geocentric distance measured in cm.  Finally, we converted all of our $Af\rho$ measurements to a common phase angle of $\alpha = 0^\circ$ using the Schleicher-Marcus phase function \citep{2010Schleicher} to give $A(0^\circ)f\rho$. The corresponding evolution of $A(0^\circ)f\rho$ vs MJD in the 1 m LCOGT $r'$-band images is shown in the middle panel of Figure~\ref{fig:aperture_dust_lc}. Visual inspection of the images used in the aperture photometry clearly revealed the presence of a surrounding dust coma. Over this time period, from UT 2025 August 10 and 2025 September 11 (MJD 60897 to MJD 60929), we measured an increase in $A(0^\circ)f\rho$ from $\sim 600$ cm to 1100 cm as the comet approached the Sun from a heliocentric distance of 3.18 au to 2.19 au. This range of $Af\rho$ measurements is consistent with other recent studies of 3I/ATLAS \citep{2025arXivScarmato} at similar heliocentric distances.

Comparing the dust production rate of 3I/ATLAS to other dynamical populations in the solar system, the dust activity is higher than what is typically observed for most JFCs at comparable heliocentric distances \citep{2024PSJGillan, 2025PSJGillan}. However, some similar activity levels have been reported for some of the more active JFCs, such as 19P/Borrelly \citep{2009IcarFink} and 67P/Churyumov–Gerasimenko \citep{2013A&ASnodgrass, 2017RSPTASnodgrass, 2022MNRASGardener, 2025PSJGillan, 2026IcarRosenbush}, particularly near perihelion. In addition, 3I/ATLAS exhibits a significantly higher dust activity than the previously discovered interstellar comet 2I/Borisov \citep{2019ApJFitzsimmons, 2019ApJJewitt}, while showing activity levels more comparable to those measured for some long-period comets at similar heliocentric distances \citep{2014A&AMazzottaEpifani, 2021P&SSGarcia}.

\subsection{The mass-loss rate}\label{subsec:mass_loss}
From $A(0^\circ)f\rho$, an approximate dust mass-loss rate, $Q_d$, in kg s$^{-1}$ can be inferred. Under simplified assumptions for the dust grain radius, ejection velocity, and grain density, we calculated an order-of-magnitude estimate of the particulate mass loss from 3I/ATLAS over the observing period \citep{2012IcarFink, 2018IcarIvanova}. However, such calculations are inherently speculative and highly sensitive to the adopted grain properties, with the inferred mass-loss rate possibly varying by at least an order of magnitude \citep{2025PSJGillan}. These assumptions are also expected to vary with heliocentric distance. We therefore calculated $Q_d$ solely to gauge the physical mass loss from the nucleus and emphasize that the resulting values should be regarded as approximate, order-of-magnitude estimates. The relationship between $Af\rho$ and the dust mass-loss rate is given in Equation~\ref{eq:Qd} by

\begin{equation}\label{eq:Qd}
Q_d = Af\rho \, \frac{4\, \pi \, r_d \, \sigma_d \, v_d}{3\, p_r}
\end{equation}

where $r_d$ is the dust grain radius, $\sigma_d$ is the grain density, $v_d$ is the dust ejection velocity, and $p_r$ is the geometric albedo in the Sloan $r'$-band. In this work, fixed dust properties were adopted; a more accurate approach would require a grain size and density distribution. Following the large-grain, low-velocity scenario proposed by \citet{2025ApJJewitt}, we assumed $r_d = 100$ $\mu$m and $v_d = 5$ m s$^{-1}$. We further adopted a typical cometary albedo of $p_r = 0.04$ and a grain density of 1 g cm$^{-3}$. To obtain the range of mass loss rates, we took the minimum $A(0^\circ)f\rho = 665 \pm 24$ cm on 2025 August 16 at 3.18 au and a maximum $A(0^\circ)f\rho = 994 \pm 40$ cm on 2025 September 11 at 2.19 au and converted these to $Q_d$. We measured an increase in the dust mass-loss rate from $\leq 217$ kg s$^{-1}$ to $\leq 328$ kg s$^{-1}$ over the $\sim1$ month observing period using the 1 m $r'$-band LCOGT data. These values represent upper limits and are comparable to mass-loss rates reported in other studies of 3I/ATLAS. \citet{2025arXivChandler} reported dust mass-loss rates of $\sim$ 10 -- 100 kg s$^{-1}$, depending on the assumed grain radius (1 -- 10 $\mu$m), while \citet{2025MNRASBolin} predicted a lower value of $\sim$ 0.1 -- 1 kg s$^{-1}$ using early, preliminary data that have since been refined. The higher mass-loss rates measured in this work are likely caused by the smaller heliocentric distance during our observations. Both comparison studies were conducted within approximately one week of discovery, where lower dust production is typically expected further from the Sun. We reiterate that all dust mass-loss estimates are highly sensitive to the underlying assumptions regarding grain properties, both here and in the comparison studies.

\subsection{The activity index}\label{subsec:AI}
We measured the activity index, $n$, defined as the power-law exponent of how $A(0^\circ)f\rho$ varies as a function of heliocentric distance, $R_{h}$, such that

\begin{equation}\label{eq:activity_index}
A(0^\circ)f\rho \propto (R_{h})^{n}
\end{equation}

\noindent The activity index was determined by fitting a power-law relation to the $A(0^\circ)f\rho$ measurements as a function of heliocentric distance, shown in the bottom panel of Figure~\ref{fig:aperture_dust_lc}. The fit was performed in log–log space using a weighted least-squares approach, with the propagated uncertainties on $A(0^\circ)f\rho$ used as weights. We measured an activity index of $n=-1.24\pm0.02$. Activity driven solely by gas sublimation, where all of the absorbed solar energy goes into sublimation, would result in an activity index of $n=-2$, due to the sublimation decreasing as the inverse square of heliocentric distance. The shallower slope we have measured therefore indicates that the dust activity of 3I/ATLAS does not scale directly with sublimation alone. Instead, $n$ is consistent with the presence of an already well-developed dust coma that evolves more gradually with decreasing heliocentric distance, rather than reflecting a sharp turn-on of activity or outbursts. This interpretation is supported by the detection of a dust coma from the time of discovery \citep{2025ApJSeligman, 2025arXivChandler}, and that it has also been active since pre-discovery at 6.5 au \citep{2025ApJYe}. In addition, the shallower slope may be influenced by the dominance of relatively slow-moving, larger dust grains \citep{2025ApJJewitt}, that take longer to accelerate and remain within the aperture for longer timescales. We also note that the activity index derived here is based on observations spanning a relatively narrow heliocentric distance range and may therefore not be representative of the comet’s long-term activity behavior. The activity index derived here is shallower than values reported in brightness-based studies of 3I/ATLAS. \citet{2025ApJJewitt} measured an activity index of $n = -3.8 \pm 0.3$ from heliocentric brightness variations over a distance range of 4.6 -- 1.8 au, while a similar brightening rate was reported in \citet{2025ApJYe}. In contrast, the index measured in this work is derived from the heliocentric variation of $A(0^\circ)f\rho$ over a narrower distance range. Since these studies measure different activity proxies, the resulting activity indices are not directly comparable. 

\section{High-cadence observations} 
Complementary to the nightly monitoring of 3I/ATLAS described in the previous sections, we obtained high-cadence intra-night imaging from two facilities: the 0.5 m UZO telescope and the D154T. These observations are densely sampled over individual nights and are therefore not included in the long-term light curve shown in Figure~\ref{fig:master_lightcurve}, which is intended to illustrate the overall magnitude evolution over the 70-day observing interval. The UZO dataset comprises 656 J--C $B$, $V$ and $R$ observations obtained over five nights between UT 2025 July 17 and 2025 July 24 (MJD 60873 -- 60880). The D154T observations were obtained over three nights at La Silla and provide densely sampled light curves in the J--C $B$, $V$, $R$, and $I$ bands. Representative single-night light curves from each facility are shown in Figure~\ref{fig:high_cadence} and the full set is presented in Appendices~\ref{uzo_append} and \ref{danish_append}, respectively. We used the densely sampled data to perform the color analysis shown in Section~\ref{subsec:colors}.

\begin{figure}
    \centering
    \subfloat[]{
        \includegraphics[width=\columnwidth]{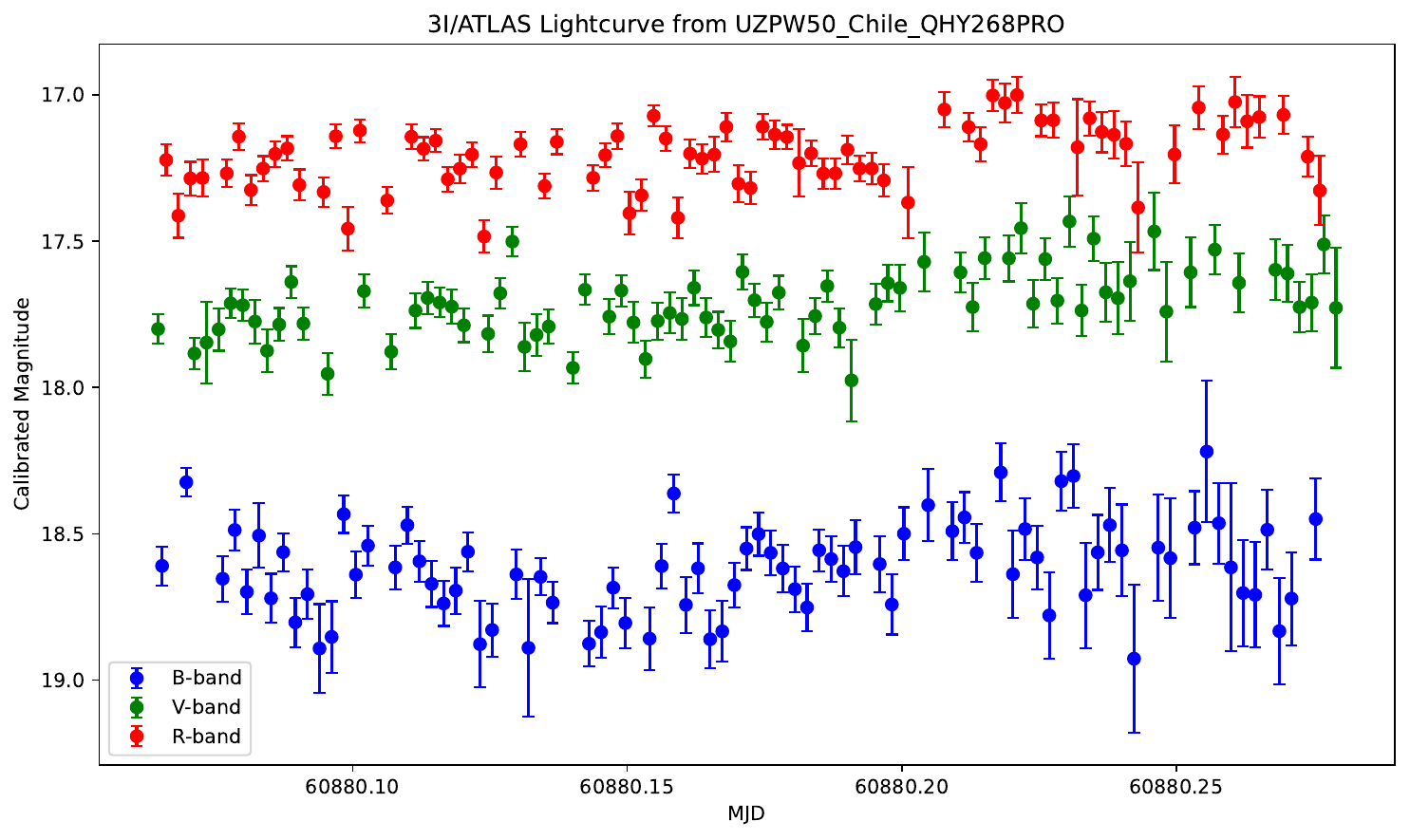}
        \label{fig:uzo_in_text}
    }\\
    \subfloat[]{
        \includegraphics[width=\columnwidth]{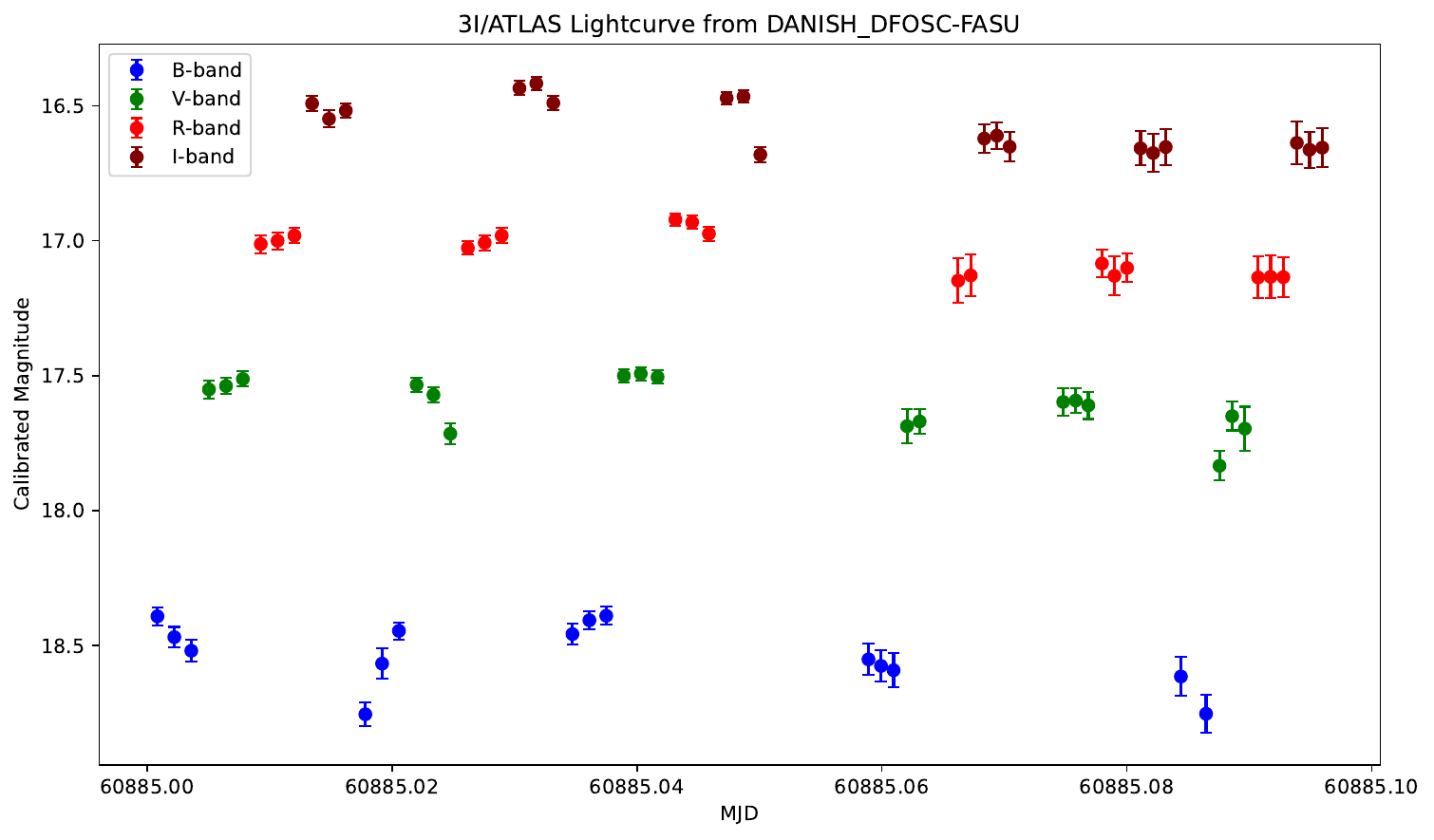}
        \label{fig:danish_in_text}
    }
    \caption{Representative high-cadence intra-night photometry of 3I/ATLAS. (a) shows UZO observations on MJD 60880 over 4.8 hours and (b) shows the D154T observations on MJD 60885 over 1.2 hours. The blue, green, red and maroon points represent the J-C $B, V, R$ and $I$ bands respectively. The remaining high-cadence light curves are shown in the Appendices~\ref{uzo_append} and \ref{danish_append}.}
    \label{fig:high_cadence}
\end{figure}

\subsection{Color analysis}\label{subsec:colors} 
In both Figure~\ref{fig:uzo_in_text} and Figure~\ref{fig:danish_in_text} there is minimal magnitude variation (less than 0.1 magnitude at most) across the J-C $B, V, R$ and $I$ bands, consistent with previous studies \citep{2025ApJKareta, 2025ApJSeligman}.
There is a wealth of color data across multiple epochs, thus we average the measured calibrated magnitudes for each filter taken on the same night for each data set. Next the D154T and UZO $BVRI$ magnitudes were converted to Sloan $g', r'$ and $i'$ magnitudes using the transformations given in \citet{lupton2005}.


The Normalized spectral reflectivity in each filter was then calculated relative to the $g'$ filter using the following equation:

\begin{equation}
R_{\lambda}=\frac{10^{-0.4[m_{\lambda}-m_{\lambda}(\odot)]}}{10^{-0.4[m_g-m_g(\odot)]}}
\end{equation}


\begin{equation}
\sigma_{R\lambda}= 0.92103\ R_{\lambda} \bigg[ \sigma_{\lambda}^{2}+\sigma_{g}^{2}+\sigma_{\lambda}(\odot)^{2}+\sigma_{g}(\odot)^{2}\bigg]^{0.5}
\end{equation}

In these relationships, $m_{\lambda}$ is the magnitude in the bandpass, $\sigma_{\lambda}$ is the uncertainty on $m_{\lambda}$, $\sigma_{r}$ is the uncertainty on $m_{r}$, and $m_{\lambda}(\odot)$ is the absolute magnitude of the Sun in that bandpass with uncertainty $\sigma_{\lambda}(\odot)$. We use $m_{g}(\odot)=5.12 \pm 0.02$, $m_{r}(\odot)=4.69 \pm 0.03$, $m_{i}(\odot)=4.57 \pm 0.03$, and $m_{z}(\odot)=4.60 \pm 0.03$.

\begin{figure*}[h!]
\centering
{\includegraphics[width=\textwidth]{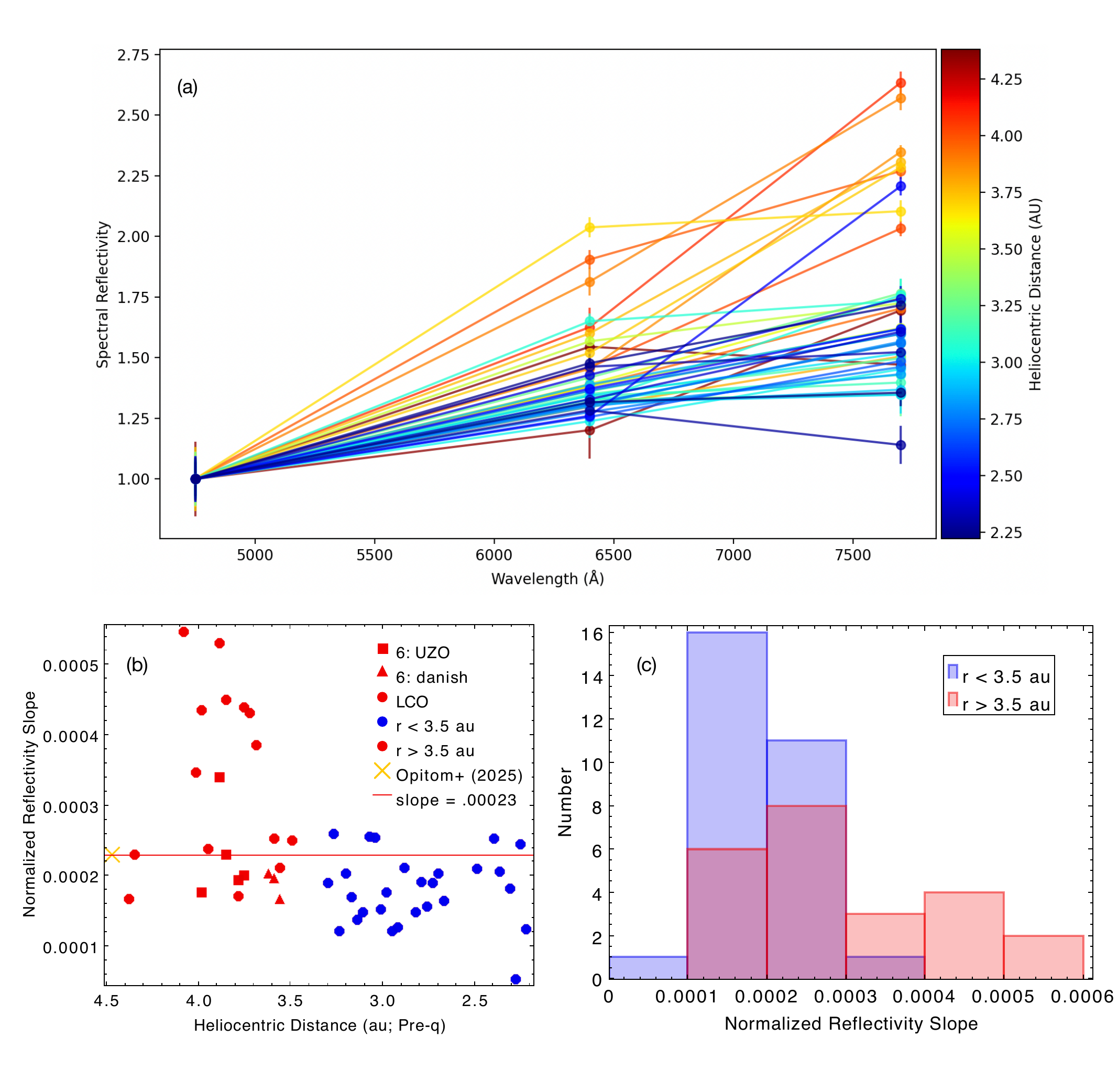}}
	\caption{}
	\label{fig:colorpanel}
\end{figure*}

Each observation's spectral reflectivity across the $g', r'$, and $i'$ filters are shown in Figure~\ref{fig:colorpanel} (a), and are color-coded by their heliocentric distance. The reddest reflectivities of 3I/ATLAS are observed when the body is further from the Sun in it's orbit, and the bluer reflectivities are highly centralized when the object is nearing 2 au.

Next we measured the spectral slope of each observation by fitting a line to the normalized reflectivities. This allowed for direct inspection of the general trend of color changes across the orbit of 3I/ATLAS. This is shown as a function of the target's pre-perihelion distance in Figure~\ref{fig:colorpanel} (b). For data taken when 3I/ATLAS was between 4.47 au and 3.5 au, we found a large scatter in colors, with a tendency towards redder colors (Figure~\ref{fig:colorpanel}c; red histogram), quantified by the large scatter in spectral slopes, and evident in the LCO lightcurve before UT 2025 August 3 (MJD 60890) in Figure~\ref{fig:master_lightcurve}. This is likely due to the variability of photometry caused by the target being near the galactic center and in a densely-crowded field. 

\begin{figure*}[h!]
\centering
{\includegraphics[width=\textwidth]{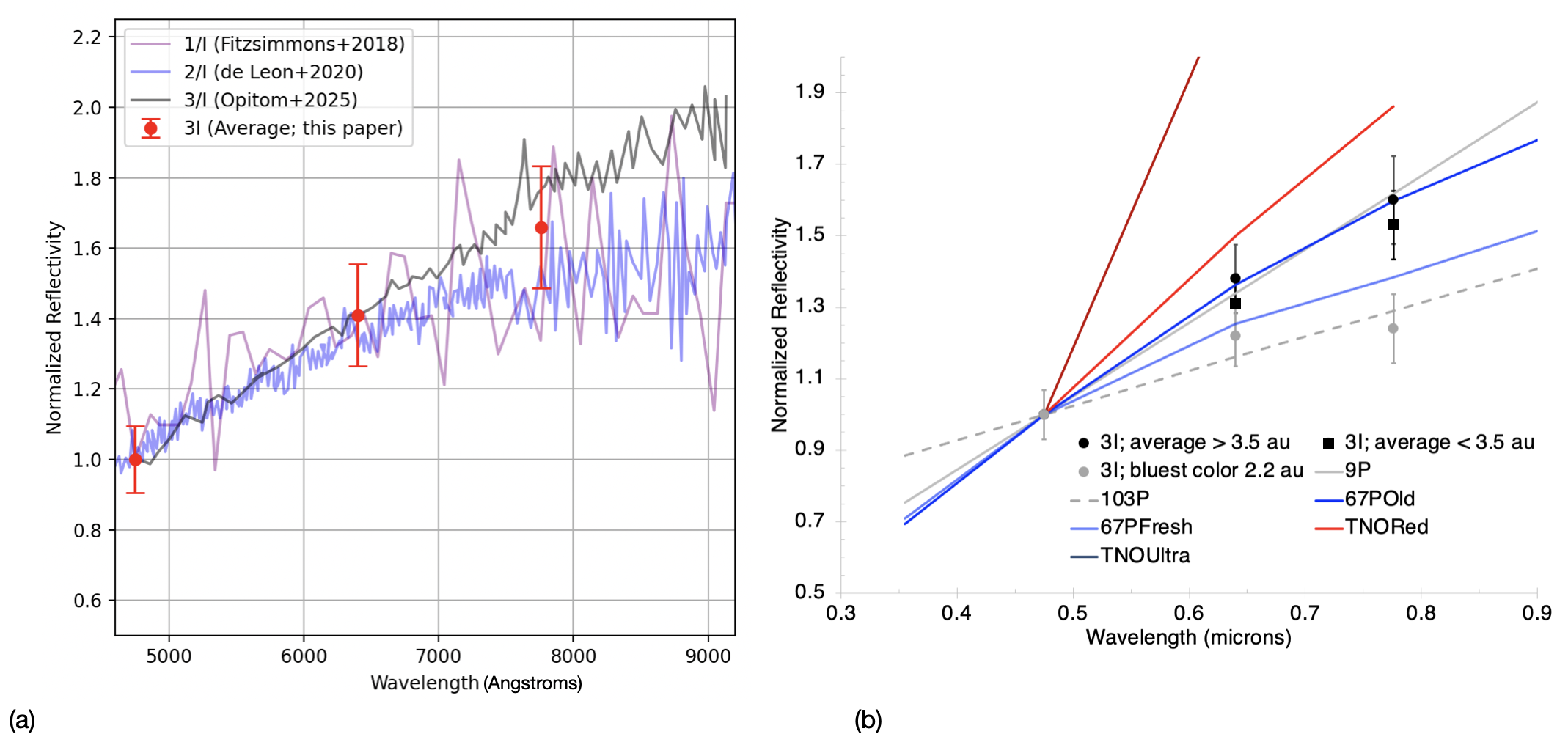}}
	\caption{}
	\label{fig:reflectivity}
\end{figure*}

The spectral slope of the normalized reflectivity measured by \citet{2025MNRASOpitom}, when the object was at 4.47 au, is displayed for comparison in Figure~\ref{fig:colorpanel}b; by a yellow cross. As this is a more robust measurement of the reflectivity than photometry, it is a reasonable assumption that the photometric colors for heliocentric distances larger than 3.5 au lie closer to this data point. Neglecting the high spectral slopes measured before 3.5 au, there may be a weaker negative trend present indicating that 3I/ATLAS is getting bluer as it approaches perihelion, however this is not statistically significant. Taking the average of all spectral slopes from the entire duration of the pre-perihelion data, it is in agreement with the spectral slope measured by \citep{2025MNRASOpitom} (Figure~\ref{fig:reflectivity}a).

We also measured the average slope pre and post 3.5 au to be $2 \cdot10^4$ and $1.8 \cdot10^4$ respectively. This minor change in slope and corresponding reflectivities is shown in comparison to other solar system small bodies in Figure~\ref{fig:reflectivity}b with black circles and squares, respectively. The maximum color difference for 3I/ATLAS, at $R_{h} < 3$ au, is also shown for comparison in Figure~\ref{fig:reflectivity}b in grey circles. If one interprets the colors as non-changing, it's spectral slope is comparable to other Jupiter-family comets such as 9P/Tempel 1 \citep{li2007,kelley2017} and even red trans-Neptunian objects (TNOs) {\citep{delsanti2004}}. If there is a real color change, it could be explained by the growing development of activity, which is seen by previous studies to have an affect on an object's color \citep{Meech2017k2, Fornasier2016, bufanda2023}. The maximum change in spectral slope possible (Figure~\ref{fig:reflectivity}b, difference between black and grey data points) would be similar to the color change observed on 67P/Churyumov-Gerasimenko (Figure~\ref{fig:reflectivity}b, blue and light blue trend) after activity revealed fresh ice on the surface of the comet \citep{Fornasier2016}. This is an upper limit on the color change; it is less likely that the effect is this pronounced because the coma of 3I/ATLAS was still developing. A color change this significant would bring 3I/ATLAS to have a spectral slope comparable to 103P/Hartley 2 \citep{li2013}. Future studies need to be performed to validate any change in spectral slope, in particular, the most robust studies will come from follow-up spectroscopy near and after perihelion.

\section{Discussion} 
\subsection{PSF and aperture photometry}
We used two independent photometric pipelines for the image processing. PSF photometry was performed using the BHTOM pipeline, calibrated to the GaiaSP catalog as described in Section~\ref{subsec:BHTOM}, while the fixed-aperture photometry was calibrated against the Gaia-DR3 catalog magnitudes (Gaia $G$, $BP$, $RP$). As magnitude measurements from both calibration methods were available in the LCOGT 1 m $r'$-band data, they were compared to assess the contribution of the extended dust coma for an active comet.

\begin{figure}
    \centering
    \includegraphics[width=\columnwidth]{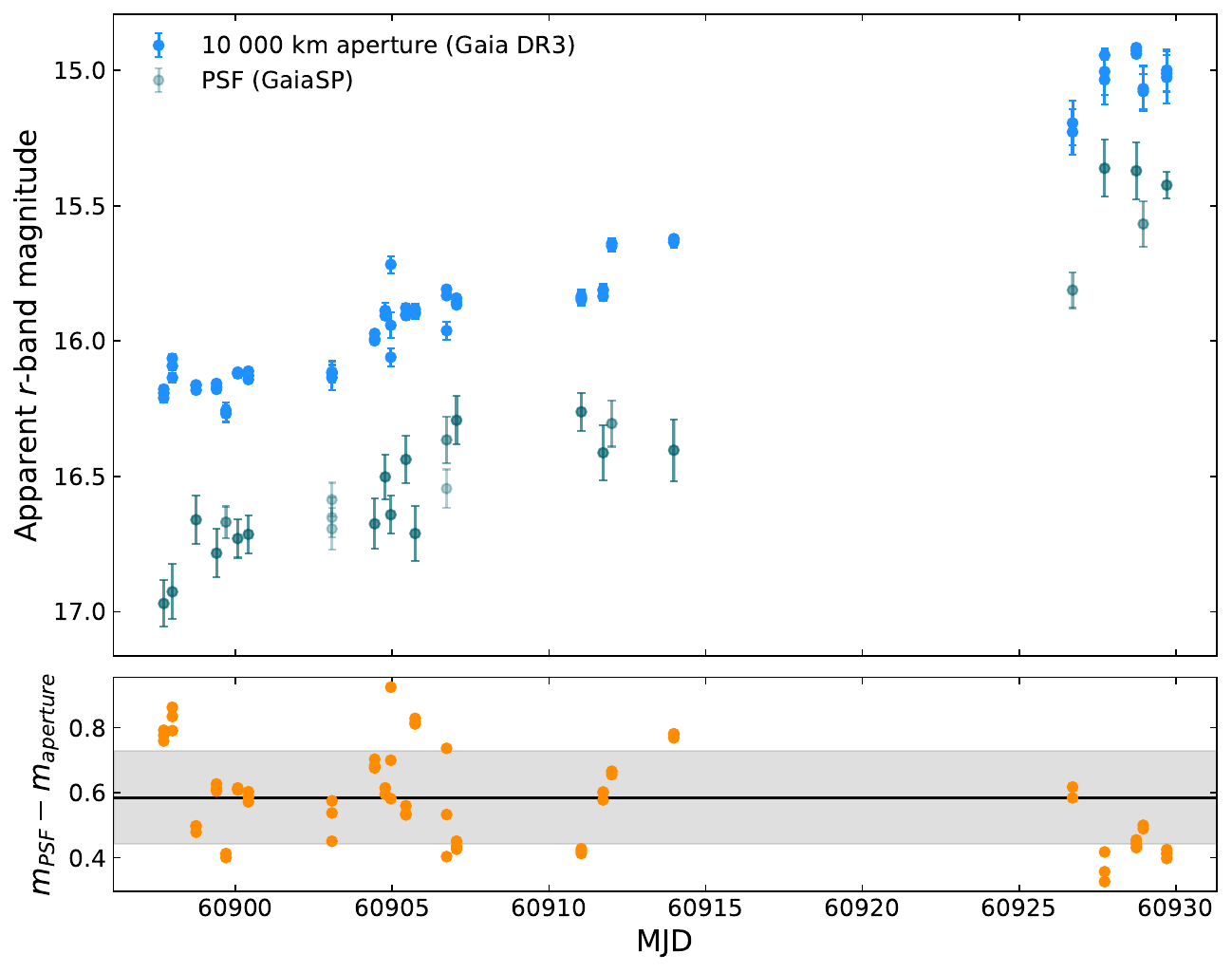}
	\caption{The upper panel shows the comparison between the PSF photometry (teal) and the fixed-aperture photometry (blue) using the LCOGT 1 m $r'$-band images. The lower panel shows the magnitude difference $m_{PSF} - m_{aperture}$ (orange) for the measurements matched in MJD, with the grey shaded region indicating the $1\sigma$ scatter from the median offset.}
	\label{fig:ap_vs_psf}
\end{figure}

During the aperture photometry analysis, measurements affected by poor image quality or aperture flux contamination were rejected as described in Section~\ref{subsec:aperture_calibrations}. As this filtering was applied specifically to the aperture photometry, not every aperture measurement had a corresponding PSF measurement obtained at the same epoch. For the comparison shown in Figure~\ref{fig:ap_vs_psf}, PSF and aperture photometry were compared on a per-image basis, and only measurements from the same exposure were included; where an exact one-to-one match was not available due to filtering, the nearest-time PSF measurement was used. The comparison is a subset of 68 matched measurements, drawn from the 83 aperture photometry points shown in Figure~\ref{fig:aperture_dust_lc}. For clarity, the PSF photometry measurements in Figure~\ref{fig:ap_vs_psf} are plotted with reduced opacity. The PSF photometry exhibits significantly less scatter than the fixed-aperture measurements, as expected, since PSF fitting is dominated by the unresolved central component, whereas aperture photometry includes flux from the extended dust coma and is more sensitive to image-to-image variations in seeing and sky subtraction. The two light curves in Figure~\ref{fig:ap_vs_psf} show the same temporal behavior but are offset by a nearly constant amount. Overall, the PSF photometry is fainter than the aperture photometry measurements, which should be expected since the aperture photometry also includes the flux from the extended dust coma. The median magnitude offset is 0.59 magnitudes which corresponds to a factor of $\sim$1.7 difference in flux. This implies that the flux measured within the fixed aperture is approximately $1.7 \times$ that which was measured using PSF photometry. Under this assumption, 40 -- 45\% of the optical flux within the aperture arises from extended dust emission rather than the nucleus.

\subsection{Implications for BHTOM in time-domain monitoring of small bodies}
The observations of 3I/ATLAS presented in this work provide a practical demonstration of coordinated, long-baseline monitoring of an active interstellar comet. The 70-day coverage, obtained using a geographically distributed network of facilities, enabled the evolution of the comet’s magnitude and dust activity to be characterized independently of short-timescale variability and survey cadence effects. Such extended, high-cadence monitoring is particularly valuable for ISOs, which can evolve rapidly during the limited intervals over which they are observable from the ground where they can become unobservable due to a lack of activity or solar elongation constraints. In this case, the sustained coverage allowed the gradual brightening and dust production of 3I/ATLAS to be observed continuously, while also supporting the interpretation of the rotational and activity-related behavior.

This study also demonstrates the use of BHTOM for coordinating non-sidereal observing campaigns across a diverse set of observing facilities. Although BHTOM was used here to provide uniform calibration and PSF photometry, users can either adopt the standard pipeline products or use BHTOM purely for coordination while applying independent photometric and calibration methods. If adopted, the pipeline delivers uniform calibrated photometry that can be used immediately to track brightness and color evolution and, where the data support it, to search for periodic signals. This flexibility is particularly relevant for small-body studies, where observing strategies and analysis can vary significantly between science cases. In addition, as BHTOM does not require traditional observing proposals, time-critical or target-of-opportunity observations can be requested from the facilities without the need for competitive allocation processes.

\section{Conclusions}
In this paper, we have presented the results from an extensive monitoring campaign of 3I/ATLAS over a 70-day observing window with data from 16 telescopes and 1554 images obtained between 2025 July 04 and 2025 September 11. Our main conclusions are as follows:

\begin{enumerate}
    \item BHTOM was used to coordinate observations and calibrate photometric data of 3I/ATLAS obtained from the observatories within the BHTOM network, enabling consistent time-domain analysis of a non-sidereal target beyond its standard application to sidereal sources and demonstrating its suitability for small-body science. Although PSF photometry is not optimal for quantifying the extended coma flux, it provided a uniform central-component light curve suitable for long-term activity trends and rotational analysis, and future small-body applications will preferentially adopt fixed-aperture photometry.
   
    \item We obtained a high-cadence, multi-band pre-perihelion light curve of 3I/ATLAS spanning 70 days, which exhibited a steady $\sim3$ magnitude increase, consistent with gradual activity evolution and showing no evidence for anomalous or outburst-like behavior.
    
    \item We measured the rotation period of 3I/ATLAS using detrended LCOGT multi-band photometry and LS periodogram analysis, adopting a rotation period of $P_{\mathrm{rot}} = 15.98 \pm 0.08$ h, consistent with values reported in previous studies.

    \item We used aperture photometry to quantify the relative dust production rate of 3I/ATLAS using the $Af\rho$ parameter, measuring an increase in $A(0^\circ)f\rho$ from $\sim 600$ to 1100 cm over MJD 60897 -- 60929 (UT 2025 August 10 and 2025 September 11) as the comet approached the Sun from 3.18 au to 2.19 au.
 
    \item Under simplified assumptions for the dust properties, we measured an increase in the dust mass-loss rate from $\leq217$ to $\leq328$ kg s$^{-1}$ over a $\sim1$ month observing period.

    \item We measured an activity index of $n=-1.24\pm0.02$, with the shallow slope indicating the presence of an already well-developed dust coma with no evidence for sudden changes in the dust production rate, consistent with the observed photometric behavior and visual inspection of the images.

    \item The colors of 3I/ATLAS are statistically non-changing. The large variance of spectral slopes between $4.47 > R_{h} > 3.5$ au was likely driven by crowded-field photometry near the Galactic center. We found a weak, non-significant tendency for 3I/ATLAS to become bluer as activity increased at $3.5 > R_{h} > 2.2$ au; however follow-up observations are required to confirm this.
\end{enumerate}
    
\begin{acknowledgements}
    A.F.G., {\L}.W. and P.P. acknowledge that this work is part of a project that has received funding from the European Union’s Horizon Europe Research and Innovation Programme under Grant Agreement No. 101131928 (ACME).
    
    {\L}.W. acknowledges support from the Polish National Science Centre DAINA grant No. 2024/52/L/ST9/00210. 

    This work makes use of observations from the Las Cumbres Observatory global telescope network.

    IST40 is one of the observational facilities of the Istanbul University Observatory, which was funded by the Scientific Research Projects Coordination Unit of Istanbul University with project numbers BAP-3685 and FBG-2017-23943.

    The data in this study were obtained with the T80 telescope at the Ankara University Astronomy and Space Sciences Research and Application Center (Kreiken Observatory) with project number of 25C.T80.07.

    TUG100 and ATA050 Telescopes at the Antalya TUG Site and Erzurum DAG Site of the Türkiye National Observatories have been utilized, and we express our gratitude for the support provided by the Türkiye National Observatories and all its staff.

    The Gravitational-wave Optical Transient Observer (GOTO) project acknowledges the support of the Monash-Warwick Alliance; University of Warwick; Monash University; University of Sheffield; University of Leicester; Armagh Observatory \& Planetarium; the National Astronomical Research Institute of Thailand (NARIT); Instituto de Astrofísica de Canarias (IAC); University of Portsmouth; University of Turku; University of Birmingham; and the UK Science and Technology Facilities Council (STFC, grant numbers ST/T007184/1, ST/T003103/1 and ST/Z000165/1).

    RFJ acknowledges the support provided by the GEMINI/ANID project under grant number 32240028, by ANID’s Millennium Science Initiative through grant ICN12\_009, awarded to the Millennium Institute of Astrophysics (MAS), and by ANID’s Basal project FB210003

    The work presented here is supported by the Carlsberg Foundation, grant CF25-0040. The grant CF25-0040 is granted to TCH/SDU-Galaxy.

    L.M. acknowledges the financial contribution from the PRIN MUR 2022 project 2022J4H55R.

    E. Khalouei has been partially supported by the National Research Foundation of Korea (NRF) grant funded by the Korea government (MSIT) (No. RS-2024-00394623).

    AM and JS acknowledge support from STFC under grant number ST/Y002563/1.
    
\end{acknowledgements}

\section*{Author Contributions}
A.F.G. led the writing and submission of the manuscript, scheduled the LCOGT observations, performed the data analysis, and led the overall project.
Ł.W. scheduled LCOGT observations, oversaw the use of BHTOM, and provided edits and improvements to the manuscript.
P.M. and K.K. were responsible for processing and calibrating the data in BHTOM.
E.B. carried out the color analysis and led the writing for the corresponding section of the manuscript.
C.O.C., H.H.H., M.S.P.K., and J.E.R. provided solar system context, interpretation and manuscript feedback. 
P.J.P. provided meaningful BHTOM and photometry discussion.
S.F. coordinated the observations from the Turkish observatories. 
R.E.C. and C.S. provided the observations from the Danish 1.54-metre Telescope.
M.Z. provided the observations from the UZO telescope.
B.H. and S.K. provided the observations from SOAB.
W.B. contributed conceptually and reviewed the manuscript
K.P. and D.O'N. led the GOTO observations. 
S.A., F.K.Y., S.Ö., M.İ., Ç.N., O.Ş., A.C.K., C.T.T. and F.T. were members of the Turkish observing teams that contributed data to this work.
M.D. was the PI on the LCOGT programme.
J.M. and M.W. were members of the BHTOM team
M.A., C.V.A., V.B., M.J.B., A.D., R.F.J., T.C.H., M.H., E.K., F.L., P.L.P., L.M., A.M., V.M., V.O., C.O., M.R., S.S., A.S.S.M., J.S., J.S., J.T.R., and R.V. were members of the MiNDSTEp consortium.
K.A., M.J.D., J.L., K.U., S.B., D.S., D.K.G., V.D., P.O'B., G.R., K.N., R.K., R.P.B., L.K.N., D.P., J.C., T.K. and A.K. were members of the GOTO collaboration.
All authors provided feedback on the manuscript.

\bibliographystyle{aa} 
\bibliography{references} 
\begin{appendix}\label{sec:appendices}
\onecolumn
\section{UZO light curves}\label{uzo_append}
\begin{figure*}[h!]
    \centering
     \resizebox{15cm}{9cm}
    {\includegraphics {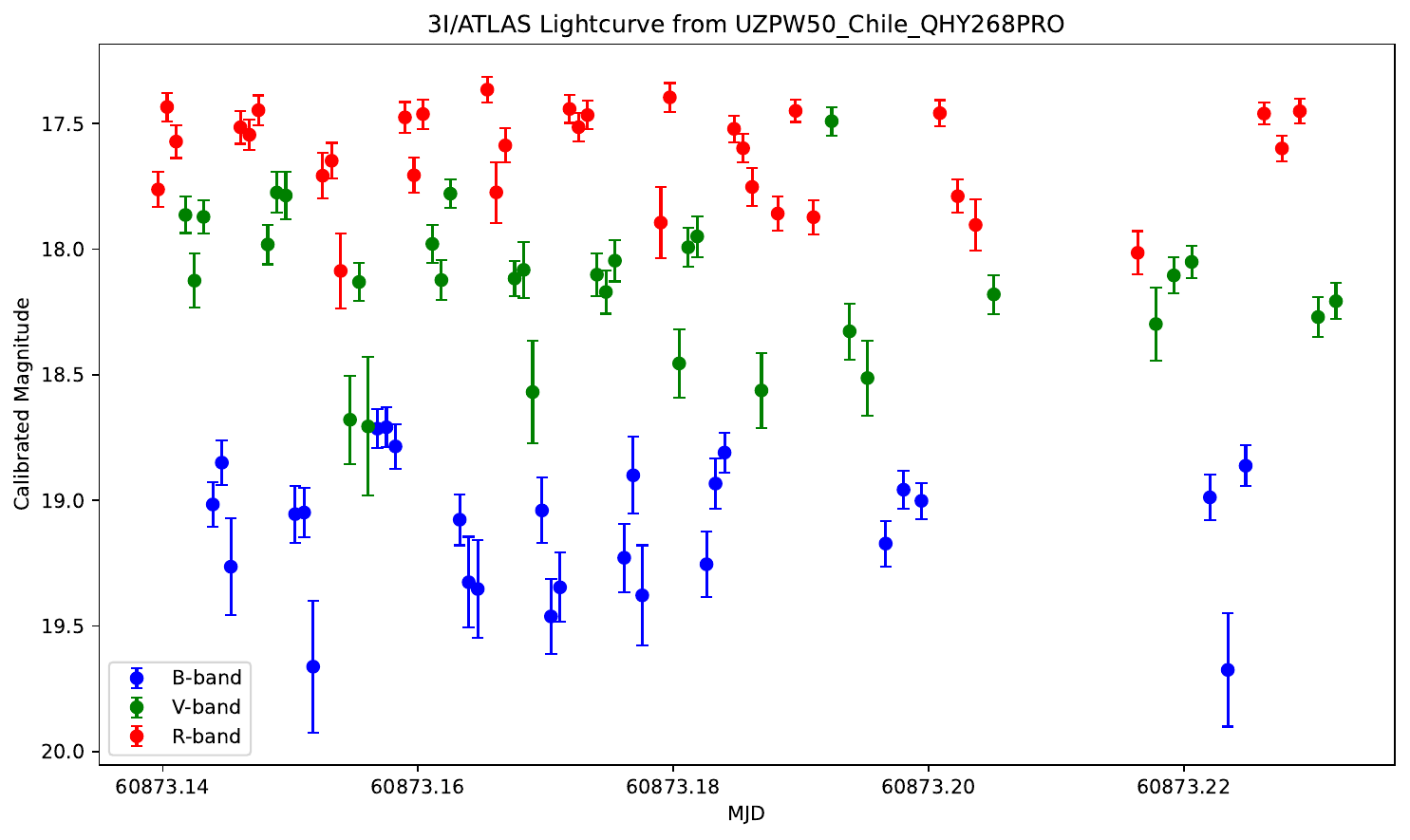}}
     \caption{The calibrated GaiaSP magnitude vs time in MJD on MJD 60873 for 2.1 hours. The blue, green and red points represent the J-C $B, V$ and $R$ bands.}
	\label{fig_apdx:uzo_1}
\end{figure*}

\begin{figure*}[h!]
    \centering
     \resizebox{15cm}{9cm}
    {\includegraphics {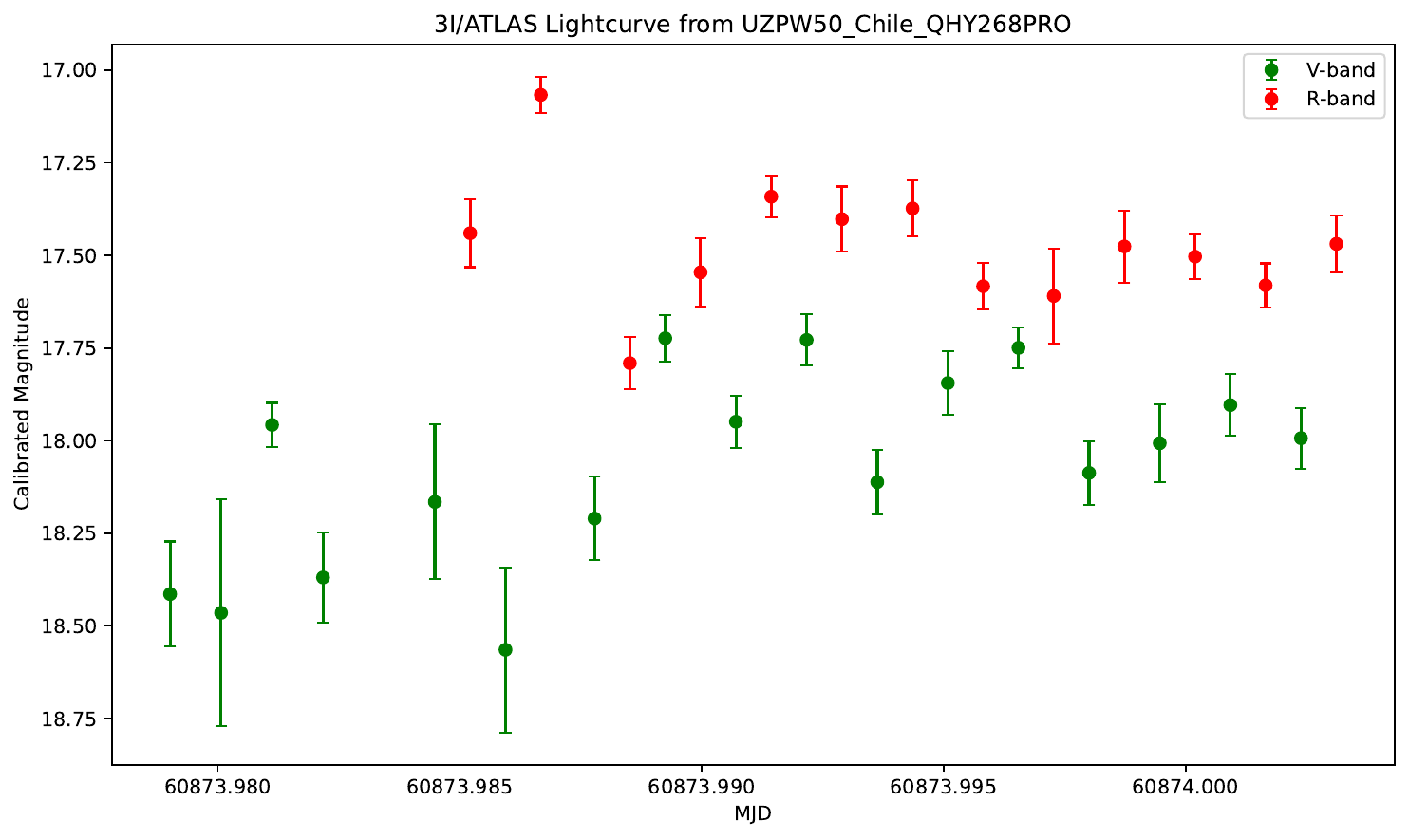}}
     \caption{The calibrated GaiaSP magnitude vs time in MJD on MJD 60873 and 60874 for 0.4 hours. The green and red points represent the J-C $V$ and $R$ bands.}
	\label{fig_apdx:uzo_2}
\end{figure*}

\begin{figure*}[h!]
    \centering
     \resizebox{15cm}{9cm}
    {\includegraphics {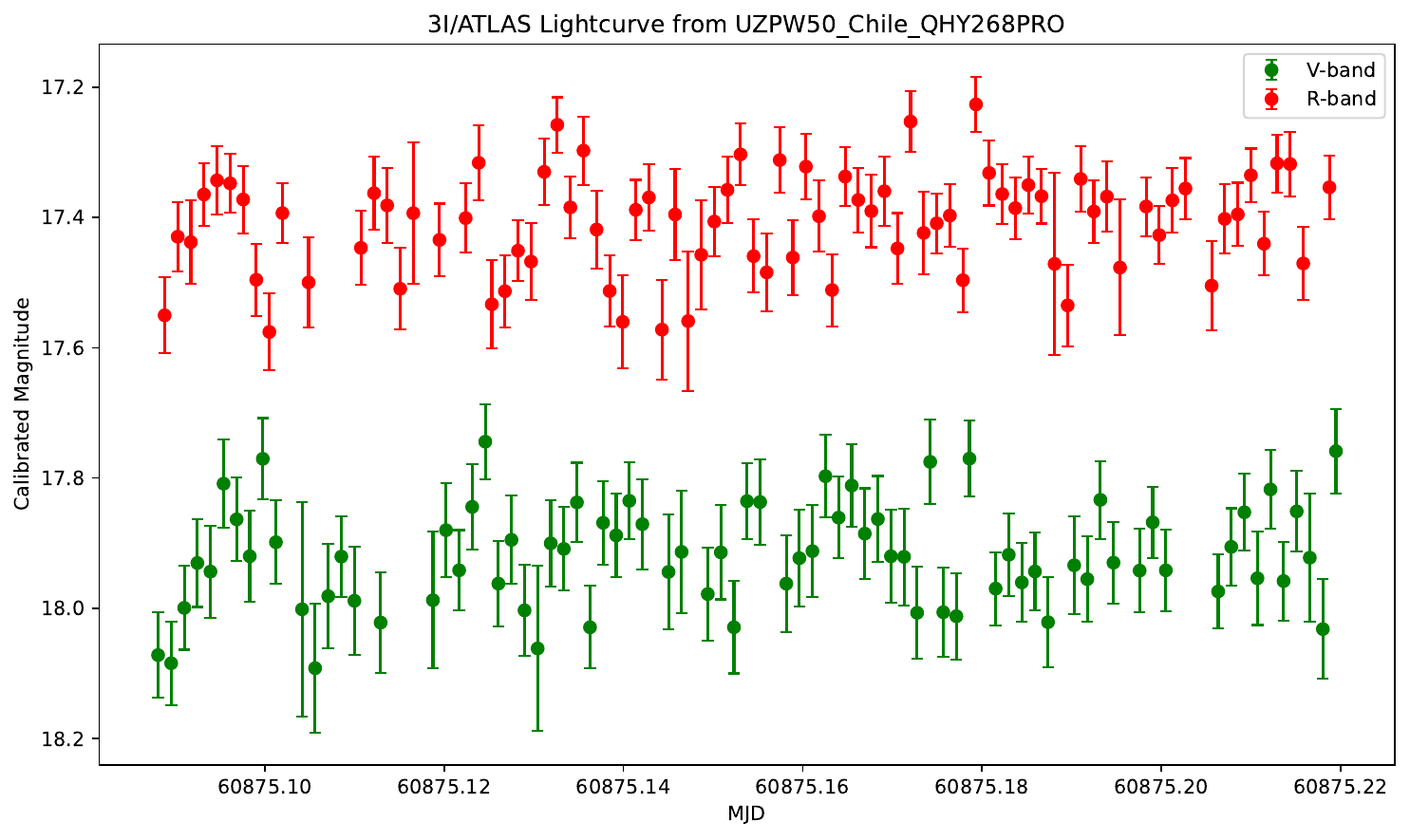}}
     \caption{The calibrated GaiaSP magnitude vs time in MJD on MJD 60875 for 2.8 hours. The blue, green and red points represent the J-C $V$ and $R$ bands.}
	\label{fig_apdx:uzo_3}
\end{figure*}

\begin{figure*}[h!]
    \centering
     \resizebox{15cm}{9cm}
    {\includegraphics {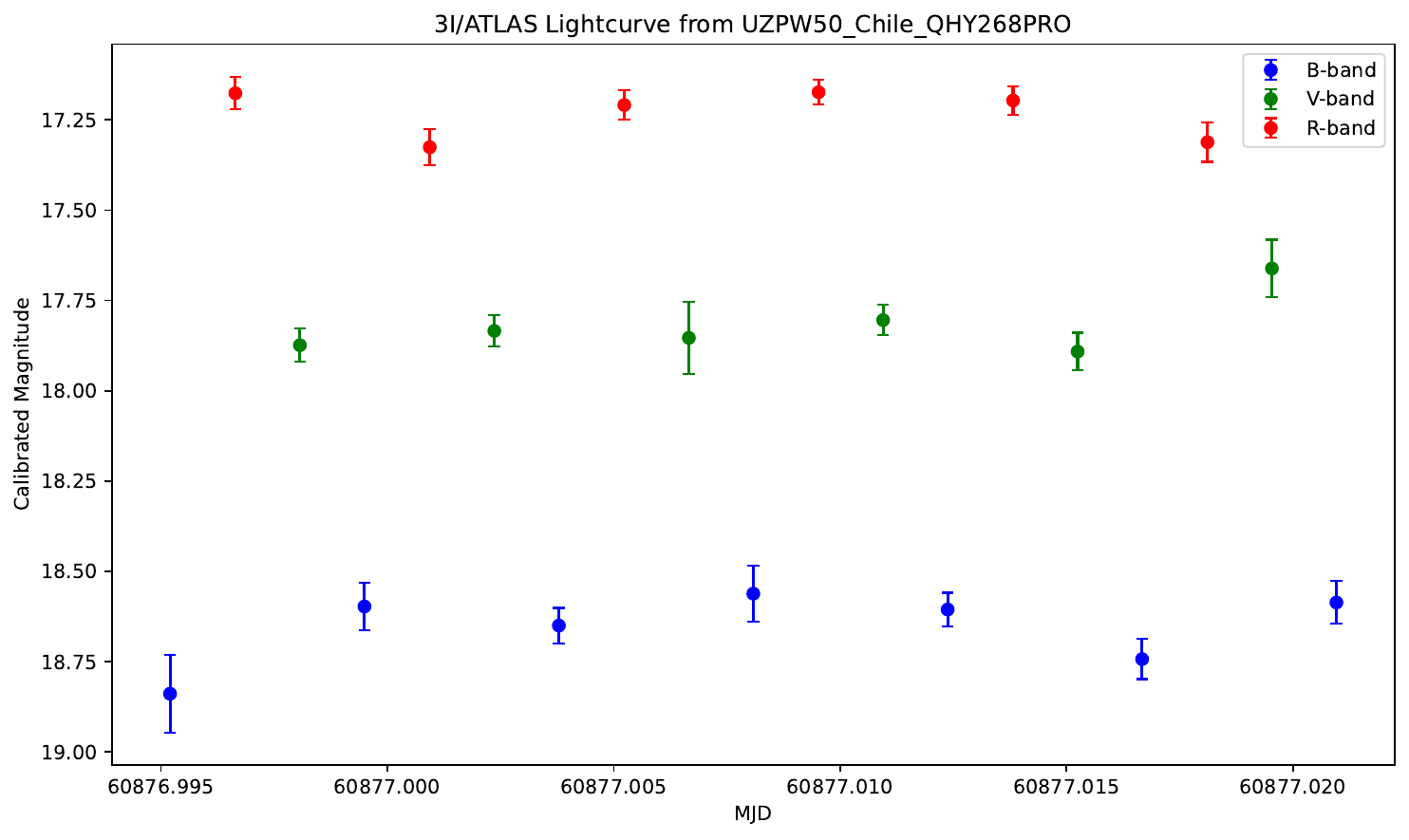}}
     \caption{The calibrated GaiaSP magnitude vs time in MJD on MJD 60876 and 60877 for 0.6 hours. The blue, green and red points represent the J-C $B, V$ and $R$ bands.}
	\label{fig_apdx:uzo_4}
\end{figure*}

\begin{figure*}[h!]
    \centering
     \resizebox{15cm}{9cm}
    {\includegraphics {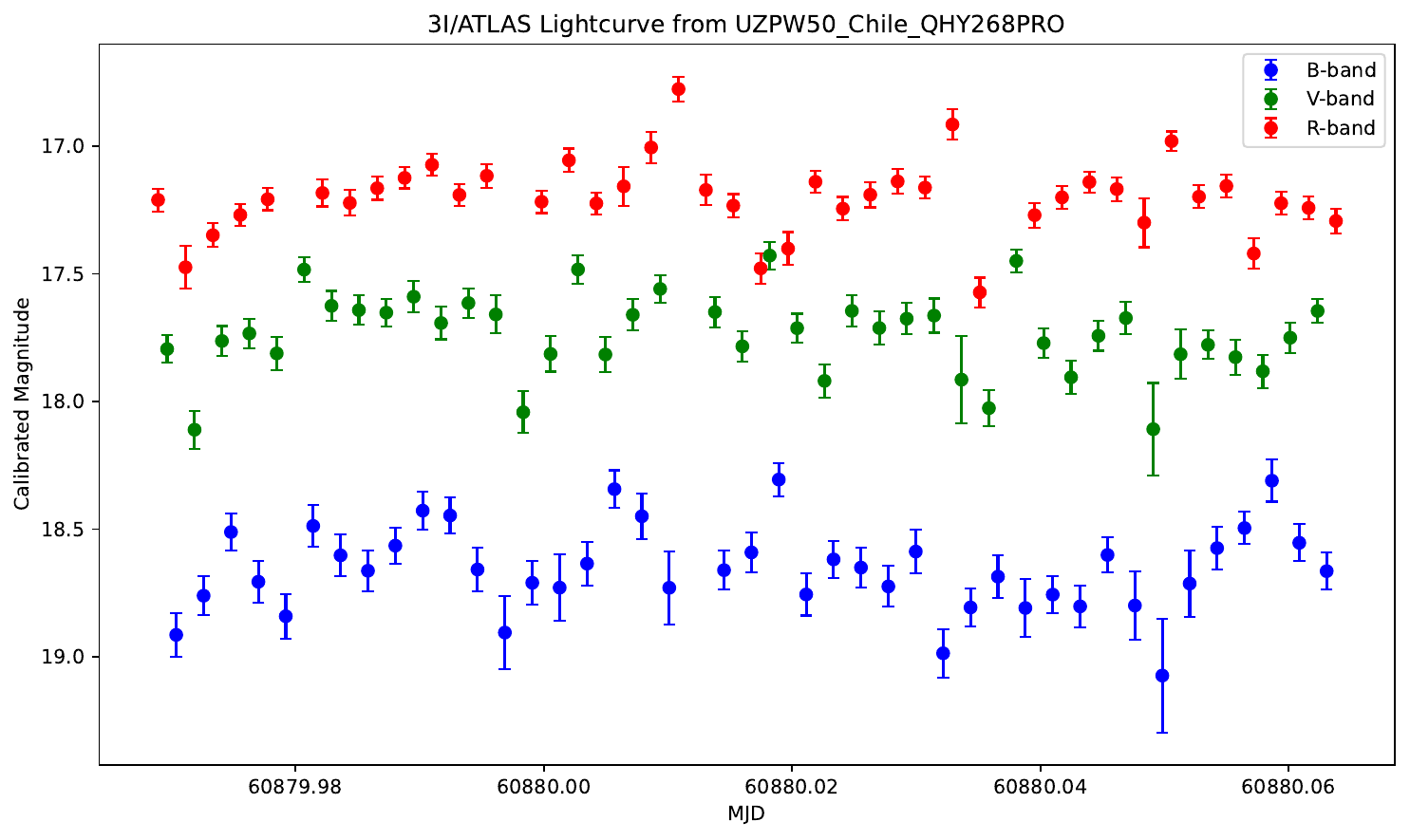}}
     \caption{The calibrated GaiaSP magnitude vs time in MJD on MJD 60879 and 60880 for 2 hours. The blue, green and red points represent the J-C $B, V$ and $R$ bands.}
	\label{fig_apdx:uzo_5}
\end{figure*}

\clearpage
\section{Danish 1.54 m Telescope light curves}\label{danish_append}

\begin{figure*}[h!]
    \centering
     \resizebox{15cm}{9cm}
    {\includegraphics {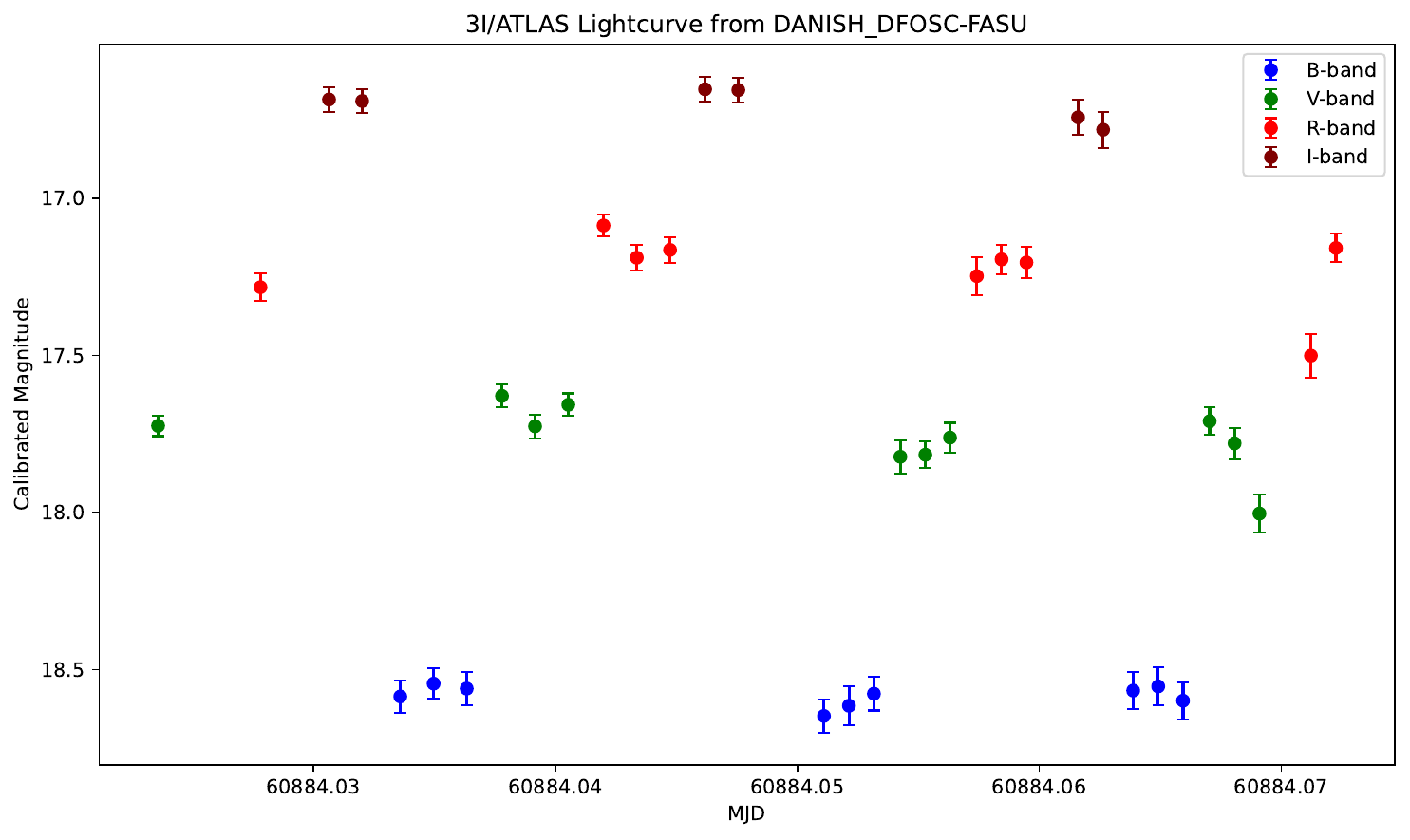}}
     \caption{The calibrated GaiaSP magnitude vs time in MJD on MJD 60884 for 0.7 hours. The blue, green, red and maroon points represent the J-C $B, V, R$ and $I$ bands.}
	\label{fig_apdx:danish_1}
\end{figure*}

\begin{figure*}[h!]
    \centering
     \resizebox{15cm}{9cm}
    {\includegraphics {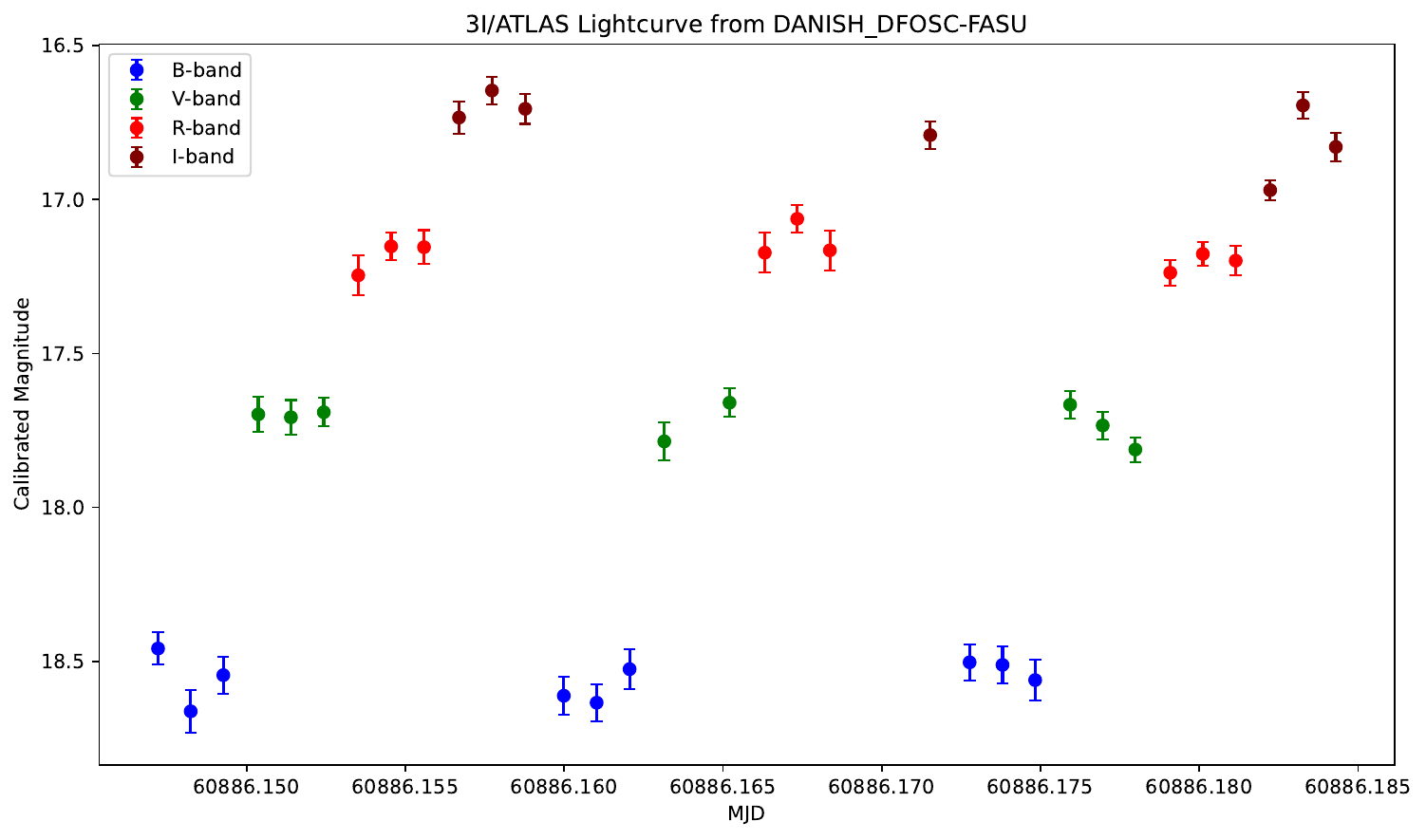}}
     \caption{The calibrated GaiaSP magnitude vs time in MJD on MJD 60886 for 0.9 hours. The blue, green, red and maroon points represent the J-C $B, V, R$ and $I$ bands.}
	\label{fig_apdx:danish_2}
\end{figure*}

\end{appendix}
\end{document}